# Running Probabilistic Programs Backwards

Neil Toronto[1] neil.toronto@gmail.com
Jay McCarthy[2] jay.mccarthy@gmail.com
David Van Horn[1] dvanhorn@cs.umd.edu

[1]University of Maryland [2]Vassar College

**Abstract.** Many probabilistic programming languages allow programs to be run under constraints in order to carry out Bayesian inference. Running programs under constraints *could* enable other uses such as rare event simulation and probabilistic verification—except that all such probabilistic languages are necessarily limited because they are defined or implemented in terms of an impoverished theory of probability. Measure-theoretic probability provides a more general foundation, but its generality makes finding computational content difficult.

We develop a measure-theoretic semantics for a first-order probabilistic language with recursion, which interprets programs as functions that compute preimages. Preimage functions are generally uncomputable, so we derive an abstract semantics. We implement the abstract semantics and use the implementation to carry out Bayesian inference, stochastic ray tracing (a rare event simulation), and probabilistic verification of floating-point error bounds.

## 1 Introduction

One key feature usually distinguishes a probabilistic programming language from general-purpose languages: finding the probabilistic conditions under which stated constraints are satisfied. Often, a probabilistic program simulates a real-world random process and the constraints represent observed, real-world outcomes. Running the program under the constraints *infers causes from effects*.

Inferring probabilistic causes from observed outcomes is called **Bayesian inference**, a technique used widely in artificial intelligence. It has been successful in analyzing phenomena at all scales, from genomes to celestial bodies. Automating it is one of the primary drivers of probabilistic language development.

One of the simplest probabilistic programs that allows us to demonstrate Bayesian inference simulates the following process of flipping two coins.

1. Flip a fair coin; call the outcome $\mathsf{x}$.
2. If $\mathsf{x}$ is heads, flip another fair coin. If $\mathsf{x}$ is tails, flip an unfair coin with heads probability 0.3 (tails probability 0.7). In either case, call the outcome $\mathsf{y}$.

The following probabilistic program simulates this process.

$$\begin{aligned}&\mathsf{let}\ \ \mathsf{x} := \mathsf{flip}\ 0.5\\&\qquad\ \mathsf{y} := \mathsf{flip}\ (\mathsf{if}\ \mathsf{x} = \mathsf{heads}\ \mathsf{then}\ 0.5\ \mathsf{else}\ 0.3)\\&\mathsf{in}\ \ \langle \mathsf{x}, \mathsf{y}\rangle\end{aligned} \tag{1}$$

Here, $\mathsf{flip}\ \mathsf{q}$ returns $\mathsf{heads}$ with probability $\mathsf{q}$ and $\mathsf{tails}$ with probability $1 - \mathsf{q}$.

The meaning of (1) is not the returned random value, but a **probability distribution** that describes the likelihoods of all possible returned random values. For discrete processes, this distribution can always be defined by a **probability mass function**: a mapping from possible values to their probabilities. These probabilities are computed by multiplying the probabilities of intermediate random values. For example, the probability of $\langle \mathsf{heads}, \mathsf{heads} \rangle$ is $0.5 \cdot 0.5 = 0.25$, and the probability of $\langle \mathsf{tails}, \mathsf{heads} \rangle$ (i.e. the second flip is unfair) is $0.5 \cdot 0.3 = 0.15$. The meaning of (1) is thus the probability mass function

$$\begin{aligned} \mathsf{p} \ :=\ & \big[\langle \mathsf{heads}, \mathsf{heads} \rangle \mapsto 0.25, \langle \mathsf{heads}, \mathsf{tails} \rangle \mapsto 0.25, \\ & \ \langle \mathsf{tails}, \mathsf{heads} \rangle \mapsto 0.15, \langle \mathsf{tails}, \mathsf{tails} \rangle \mapsto 0.35\big] \end{aligned} \tag{2}$$

Using $\mathsf{p}$, we can answer any question about the process under constraints. For example, if we do not know $\mathsf{x}$, but constrain $\mathsf{y}$ to be $\mathsf{heads}$, what is the probability that $\mathsf{x}$ is also $\mathsf{heads}$? We compute the answer by dividing the probability of the outcome we are interested in (i.e. $\langle \mathsf{x}, \mathsf{y} \rangle = \langle \mathsf{heads}, \mathsf{heads} \rangle$) by the total probability of outcomes in the constraint's corresponding subdomain $\{\mathsf{heads}, \mathsf{tails}\} \times \{\mathsf{heads}\}$:

$$\frac{\mathsf{p}\ \langle \mathsf{heads}, \mathsf{heads} \rangle}{\sum_{\mathsf{z} \in \{\mathsf{heads}, \mathsf{tails}\} \times \{\mathsf{heads}\}} \mathsf{p}\ \mathsf{z}} \ =\ \frac{0.25}{0.25 + 0.15} \ =\ 0.625 \tag{3}$$

Qualitatively, $\mathsf{y}$ being $\mathsf{heads}$ is a bit unusual if the second coin is unfair. Therefore, we infer that the second coin is most probably fair; i.e. $\mathsf{x}$ is most likely $\mathsf{heads}$.

The time complexity of computing $\mathsf{p}$ is generally exponential in the number of random choices, which is intractable for all but the simplest processes. One popular way to avoid this exponential explosion is to use advanced Monte Carlo algorithms to sample according to $\mathsf{p}$ on the constraint's corresponding subdomain without explicitly enumerating that subdomain. The number of samples required is typically quadratic in the answer's desired accuracy [7, Sec. 12.2].

Probabilistic languages that are implemented using advanced Monte Carlo algorithms could be used not just for Bayesian inference, but for simulating **rare events** (i.e. very low-probability events) by encoding the events as constraints.

Stochastic ray tracing [30] is one such rare-event simulation task. As illustrated in Fig. 1, to carry out stochastic ray tracing, a probabilistic program simulates a light source emitting a single photon in a random direction, which is reflected or absorbed when it hits a wall. The program outputs the photon's path, which is constrained to pass through an aperture. Millions of paths that meet the constraint are sampled, then projected onto a simulated sensor array.

The program's main loop is a recursive function with two arguments: $\mathsf{path}$, the photon's path so far as a list of points, and $\mathsf{dir}$, the photon's current direction.

$$\begin{array}{l} \mathsf{simulate\text{-}photon\ path\ dir}\ := \\ \quad \mathsf{case}\ \ (\mathsf{find\text{-}hit}\ (\mathsf{fst\ path})\ \mathsf{dir})\ \ \mathsf{of} \\ \qquad \mathsf{absorb\ pt} \qquad\quad \longrightarrow\ \langle \mathsf{pt}, \mathsf{path} \rangle \\ \qquad \mathsf{reflect\ pt\ norm}\ \ \longrightarrow\ \mathsf{simulate\text{-}photon}\ \langle \mathsf{pt}, \mathsf{path} \rangle\ (\mathsf{random\text{-}half\text{-}dir\ norm}) \end{array} \tag{4}$$

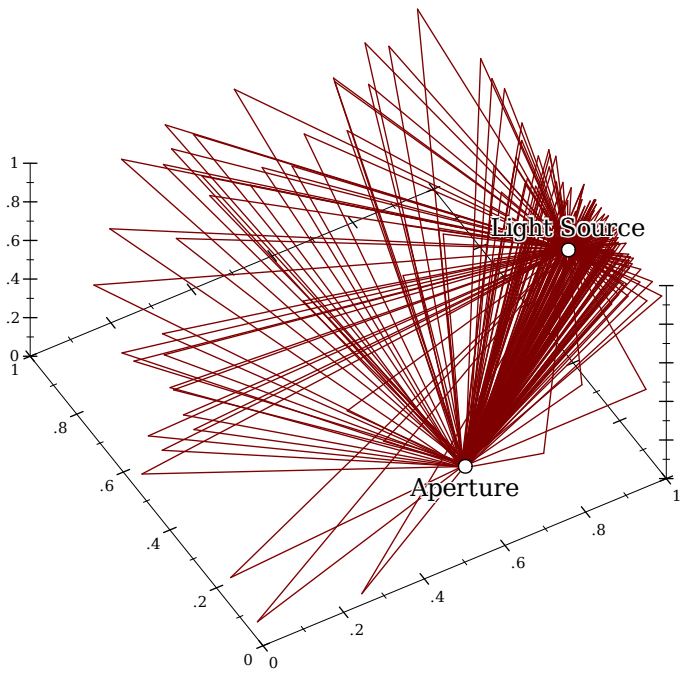

(a) Simulated photons from a single source, constrained to pass through an aperture.

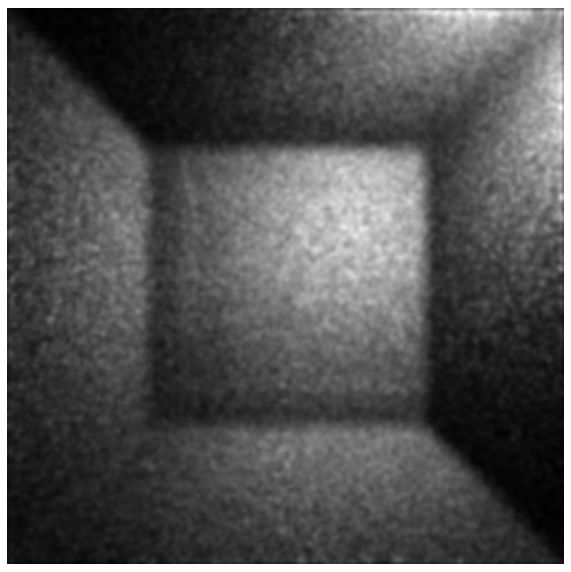

(b) Simulated photons constrained to pass through the aperture, projected onto a plane and accumulated.

Fig. 1: Ray tracing by constraining the outputs of a probabilistic program.

Here, find-hit (fst path) dir finds the surface the photon hits. If the photon is absorbed, find-hit returns a data structure containing just the collision point pt. Otherwise, find-hit returns a data structure containing the collision point pt and surface normal norm, which random-half-dir uses to choose a new direction. Running simulate-photon $\langle$pt, $\langle\rangle\rangle$ dir, where pt is the light source's location and dir is a random emission direction, generates a photon path. The fst of the path (the last collision point) is constrained to be in the aperture. The remainder of the program is simple vector math that computes ray-plane intersections.

In contrast, hand-coded stochastic ray tracers, written in general-purpose languages, are much more complex and divorced from the physical processes they simulate, because they must interleave the advanced Monte Carlo algorithms that ensure the aperture constraint is met.

Unfortunately, while many probabilistic programming languages support random real numbers, none are capable of running a probabilistic program like (4) under constraints to carry out stochastic ray tracing. The reason is not lack of engineering or weak algorithms, but is theoretical at its core: they are all either defined or implemented using a naive theory of probability.

While probability mass functions cannot define distributions on $\mathbb{R}$ that give positive probability to uncountably many values, there is a near-universal substitute that can: probability *density* functions. Density functions map single values to probability-like quantities, which makes them intuitively appealing and apparently simple. Unfortunately, density functions are not general enough to be used as probabilistic program meanings without imposing severe limitations on probabilistic languages. In particular, programs whose outputs are deterministic functions of random values and programs with recursion generally cannot denote density functions. The program in (4) exhibits both characteristics.

Measure-theoretic probability is a more powerful alternative to this naive probability theory based on probability mass and density functions. It not only subsumes naive probability theory, but is capable of defining any computable probability distribution, and many uncomputable distributions. But while even the earliest work [15] on probabilistic languages is measure-theoretic, the theory's generality has historically made finding useful computational content difficult.

We show that measure-theoretic probability can be made computational by

1. Using measure-theoretic probability to define a compositional, denotational semantics that gives a valid denotation to every program.
2. Deriving an abstract semantics, which allows computing answers to questions about probabilistic programs to arbitrary accuracy.
3. Implementing the abstract semantics and efficiently solving problems.

In fact, our primary implementation, *Dr. Bayes*, produced Fig. 1b by running a probabilistic program like (4) under an aperture constraint.

The rest of this paper is organized as follows.

- Section 2 demonstrates why density functions are insufficient for interpreting probabilistic programs. It shows how measure-theoretic probability defines probability distributions using set-valued inverses, or *preimage functions*.
- Section 3 presents the categorical tools we use to derive many semantics from a single standard semantics in a way that makes them easy to prove correct.
- Section 4 defines the semantics of nonrecursive, nonprobabilistic programs, which interprets programs as preimage functions.
- Section 5 lifts this semantics to recursive, probabilistic programs.
- Section 6 derives a sound, implementable abstract semantics.
- Section 7 describes our implementations and gives examples, including probabilistic verification of floating-point error bounds.

In short, we show why and how to run probabilistic programs under constraints by computing preimage functions—that is, by running programs backwards.

## 2 Background

### 2.1 Probability Density Functions

Some distributions of real values can be defined by **probability density functions**: integrable functions $\mathsf{p} : \mathbb{R}^{\mathsf{n}} \to [0, \infty)$ that integrate to 1.

The simplest nontrivial probabilistic program is $\mathsf{random}$, which returns a uniformly random value in the interval $[0, 1]$. The meaning of $\mathsf{random}$ is a probability distribution that can be defined by the density

$$\mathsf{p} : \mathbb{R} \to [0, \infty) \qquad \mathsf{p}\ \mathsf{x}\ :=\ \begin{cases} 1 & \text{if } \mathsf{x} \in [0,1] \\ 0 & \text{otherwise} \end{cases} \tag{5}$$

Though $\mathsf{p}\ \mathsf{x}$ for any $\mathsf{x}$ indicates $\mathsf{x}$'s relative frequency, $\mathsf{p}\ \mathsf{x}$ is not a probability. Probabilities are obtained by integration. For example, the probability that $\mathsf{random}$ returns a value in $[0, 0.5]$ is

$$\int_0^{0.5} (\mathsf{p}\ \mathsf{x})\, d\mathsf{x}\ =\ \int_0^{0.5} 1\ d\mathsf{x}\ =\ \Big[\mathsf{x}\Big]_0^{0.5}\ =\ 0.5 - 0\ =\ 0.5 \tag{6}$$

Similarly, the probability of $[0.5, 0.5]$ or any other singleton set is zero. In fact, *every* probability density function integrates to zero on singleton sets.

This fact makes it trivial to write a probabilistic program whose distribution cannot be defined by a density. For example, consider $\mathsf{max}\ \langle 0.5, \mathsf{random} \rangle$, where $\mathsf{max}\ \langle \mathsf{a}, \mathsf{b} \rangle$ returns the greater of the pair $\langle \mathsf{a}, \mathsf{b} \rangle$. This program evaluates to 0.5 whenever $\mathsf{random}$ returns a number in $[0, 0.5]$. In other words, the value of $\mathsf{max}\ \langle 0.5, \mathsf{random} \rangle$ is in $[0.5, 0.5]$ with probability 0.5. But if its distribution is defined by a density, then $[0.5, 0.5]$ must have probability zero—not 0.5.

A probabilistic language without the $\mathsf{max}$ function can still be useful. It is fairly easy to compute densities for the outputs of single-argument functions that happen to have differentiable inverses, such as exponentiation and square root. But two-argument functions such as addition and multiplication require evaluating integrals, which generally do not have closed-form solutions.

Perhaps the most constricting limitation of probability density functions is that the number of dimensions must be finite and fixed. This limitation rules out recursive data types, and makes recursion so difficult that few probabilistic languages attempt to allow it.

### 2.2 Measures, and Measures of Preimages

Measure-theoretic probability gains its expressive power by mapping sets directly to probabilities. Functions that do so are called **probability measures**. For example, the distribution of $\mathsf{random}$ is defined by the probability measure

$$\mathsf{P} : \mathcal{P}\ [0,1] \rightharpoonup [0,1] \qquad \mathsf{P}\ [\mathsf{a}, \mathsf{b}] \ = \ \mathsf{b} - \mathsf{a} \tag{7}$$

where $\mathcal{P}\ [0,1]$ is the powerset of $[0,1]$ and '$\rightharpoonup$' denotes a partial mapping. Though (7) apparently defines $\mathsf{P}$ only on intervals, it is regarded as defining $\mathsf{P}$ additionally on countable unions of intervals, their complements, countable unions of such, and so on. The resulting domain includes almost every subset of $[0,1]$ that can be written down.

Probability measures can be defined on any domain, including domains with variable and infinite dimension. They can also map singleton sets to nonzero probabilities, which we will demonstrate shortly by deriving a probability measure for $\mathsf{max}\ \langle 0.5, \mathsf{random} \rangle$.

Measure-theoretic probability takes great pains to separate random effects from the pure logic of mathematics. It does so in the same way Haskell and other purely functional programming languages allow random effects: by interpreting probabilistic processes as *deterministic functions* that operate on an assumed-random source. The probabilities of sets of outputs are uniquely determined by the probabilities of the corresponding sets of inputs.

Suppose we interpret $\mathsf{max}\ \langle 0.5, \mathsf{random} \rangle$ as the deterministic function

$$\mathsf{f} \ := \ \lambda\, \mathsf{r} \in [0,1].\, \mathsf{max}\ \langle 0.5, \mathsf{r} \rangle \tag{8}$$

and assume that $\mathsf{r}$ is its uniform random source; i.e. that its distribution is $\mathsf{P}$ as defined in (7). To compute the probability that $\mathsf{max}\ \langle 0.5, \mathsf{random} \rangle$ evaluates to 0.5, we apply $\mathsf{P}$ to the set of all $\mathsf{r}$ for which $\mathsf{f\ r} \in [0.5, 0.5]$, and get, as expected,

$$\mathsf{P}\ \{\mathsf{r} \in [0,1] \mid \mathsf{f\ r} \in [0.5, 0.5]\} \;=\; \mathsf{P}\ [0, 0.5] \;=\; 0.5 - 0 \;=\; 0.5 \tag{9}$$

For any $\mathsf{f}$ and $\mathsf{B}$, the set $\{\mathsf{a} \in \mathsf{domain\ f} \mid \mathsf{f\ a} \in \mathsf{B}\}$ is called the **preimage of** $\mathsf{B}$ **under** $\mathsf{f}$. Functions that compute preimages are often denoted $\mathsf{f}^{-1}$ to emphasize that they are a sort of generalized inverse function. However, we find this notation confusing: inverse functions operate on *values* and may not be well-defined, whereas preimage functions operate on *sets* and are *always* well-defined.[1] Thus, we denote $\mathsf{f}$'s preimage function by $\mathsf{preimage\ f}$. The probability that $\mathsf{f}$ outputs a value in $\mathsf{B}$ is therefore $\mathsf{P}\ ((\mathsf{preimage\ f})\ \mathsf{B})$, or $\mathsf{P}\ (\mathsf{preimage\ f\ B})$.

Though the distribution of $\mathsf{max}\ \langle 0.5, \mathsf{random}\rangle$, or the output of $\mathsf{f}$, has no probability density function, its probability measure is defined by

$$\mathsf{P_f} : \mathcal{P}\ [0.5, 1] \rightharpoonup [0,1] \qquad \mathsf{P_f}\ [\mathsf{a}, \mathsf{b}] \;=\; \mathsf{P}\ (\mathsf{preimage\ f}\ [\mathsf{a}, \mathsf{b}]) \tag{10}$$

An equivalent, more elegant definition is

$$\mathsf{P_f} \;:=\; \mathsf{P} \circ (\mathsf{preimage\ f}) \tag{11}$$

which clearly shows that $\mathsf{P_f}$ is factored into a part $\mathsf{P}$ that quantifies randomness, and a deterministic part $\mathsf{preimage\ f}$ that *runs* $\mathsf{f}$ *backwards on sets of outputs.*

This factorization confers the flexibility to interpret probabilistic programs by choosing any $\mathsf{P}$ and $\mathsf{f}$ for which $\mathsf{P} \circ (\mathsf{preimage\ f})$ is the correct measure. For $\mathsf{P}$, we choose uniform measures on cartesian products of $[0,1]$ (e.g. $[0,1]^{\mathbb{N}}$) and interpret each $\mathsf{random}$ as a projection. Thus, for the remainder of this paper, we can concentrate solely on computing $\mathsf{preimage\ f}$.

Because $\mathsf{preimage\ f}$ is deterministic, techniques to compute it have applications outside of probabilistic programming; for example, constraint-functional languages, type inference, and verification. More immediately, its determinism means that, for the bulk of this paper, *readers do not need to know anything about probability, let alone measure theory*—only basic set theory.

### 2.3 Preimage Semantics

Several well-known identities suggest that preimages can be computed compositionally, which would make it possible to define a denotational semantics that interprets programs as preimage functions. For example, we have

$$\begin{aligned} \mathsf{preimage\ id} &= \mathsf{id} \\ \mathsf{preimage}\ (\mathsf{f_2} \circ \mathsf{f_1}) &= (\mathsf{preimage\ f_1}) \circ (\mathsf{preimage\ f_2}) \\ \mathsf{preimage}\ \langle \mathsf{f_1}, \mathsf{f_2}\rangle\ (\mathsf{B_1} \times \mathsf{B_2}) &= (\mathsf{preimage\ f_1\ B_1}) \cap (\mathsf{preimage\ f_2\ B_2}) \end{aligned} \tag{12}$$

where $\langle \mathsf{f_1}, \mathsf{f_2}\rangle = \lambda\, \mathsf{a} \in (\mathsf{domain\ f_1}) \cap (\mathsf{domain\ f_2}).\ \langle \mathsf{f_1\ a}, \mathsf{f_2\ a}\rangle$ constructs pairing functions and $\mathsf{id}$ is the identity function.

It might seem we can easily use identities like those in (12) directly to define a semantic function $[\![\cdot]\!]_{\mathsf{pre}}$ that interprets programs as preimage functions. Unfortunately, our task is not that simple, for the following reasons.

[1] If $\mathsf{f}^{-1}\ \mathsf{b}$ is undefined, then the preimage of $\{\mathsf{b}\}$ under $\mathsf{f}$ is simply $\varnothing$.

1. The $\mathsf{preimage}$ function requires its argument to have an observable domain. This includes **extensional** functions, which are sets of intput/output pairs (i.e. possibly infinite hash tables), but not **intensional** functions, which are syntactic rules for computing outputs from inputs (e.g. lambdas).[2]
2. We must ensure $\mathsf{preimage}\ \mathsf{f}\ \mathsf{B}$ is always in the domain of the chosen probability measure $\mathsf{P}$. (Recall that probability measures are partial functions.) If this is true, we say $\mathsf{f}$ is **measurable**. Proving measurability is difficult, especially if $\mathsf{f}$ may not terminate.
3. The function $\mathsf{app} : (\mathsf{X} \to \mathsf{Y}) \times \mathsf{X} \to \mathsf{Y}$, when restricted to measurable functions, is not generally measurable if we want good approximation properties [2]. This makes interpreting higher-order application difficult.

Implementing a language based on preimage semantics is complicated because

4. Ordinary set-based mathematics is unlike any implementation language.
5. It requires running programs written in a Turing-equivalent language backwards, efficiently, on possibly uncountable sets of outputs.

We address 1 and 4 by developing our semantics using $\lambda_{\mathrm{ZFC}}$ [29], an untyped, call-by-value $\lambda$-calculus with infinite sets, real numbers, extensional functions such as $\lambda\, \mathsf{r} \in [0,1].\, \mathsf{max}\ \langle 0.5, \mathsf{r}\rangle$, intensional functions such as $\lambda\, \mathsf{r}.\, \mathsf{max}\ \langle 0.5, \mathsf{r}\rangle$, a computable sublanguage, and an operational semantics. It is essentially ordinary mathematics extended with lambdas and general recursion, or equivalently a lambda calculus extended with uncountably infinite sets and set operations.

We have addressed difficulty 2 by proving that all programs' interpretations as functions are measurable if language primitives are measurable, including uncomputable primitives such as limits and real equality, regardless of nontermination. The proof interprets programs as extensional functions and applies well-known theorems from measure theory such as the identities in (12). Unfortunately, the required machinery does not fit in this paper; see the first author's dissertation [28] for the entire development.

We avoid difficulty 3 for now by interpreting a language with *first-order* functions and recursion. We address 5 by deriving and implementing a *conservative approximation* of the preimage semantics, and using its approximations to compute measures of preimages with arbitrary accuracy.

### 2.4 Abstract Interpretation, Categorically

We interpret nonrecursive, nonprobabilistic programs three different ways, using

1. A **standard semantics** $[\![\cdot]\!]_\perp$ that interprets programs that may raise errors (e.g. divide-by-zero) as intensional functions.
2. A **concrete semantics** $[\![\cdot]\!]_{\mathsf{pre}}$ that interprets programs as preimage functions, which operate on uncountable sets, and are thus unimplementable.
3. An **abstract semantics** $[\![\cdot]\!]_{\widehat{\mathsf{pre}}}$ that interprets programs as *abstract* preimage functions, which operate only on overapproximating, finite representations of uncountable sets, and thus *are* implementable.

[2] The lambda $\lambda\, \mathsf{r}.\, \mathsf{max}\ \langle 0.5, \mathsf{r}\rangle$ is intensional, but $\lambda\, \mathsf{r} \in [0,1].\, \mathsf{max}\ \langle 0.5, \mathsf{r}\rangle$ constructs an extensional function by pairing every $\mathsf{r} \in [0,1]$ with its corresponding $\mathsf{max}\ \langle 0.5, \mathsf{r}\rangle$.

Of course, we must prove for any program $p$, that $[\![p]\!]_{\mathsf{pre}}$ correctly computes preimages under $[\![p]\!]_{\bot}$, and that $[\![p]\!]_{\widehat{\mathsf{pre}}}$ is sound with respect to $[\![p]\!]_{\mathsf{pre}}$.

For recursive, probabilistic programs, we define three more semantic functions analogous to $[\![\cdot]\!]_{\bot}$, $[\![\cdot]\!]_{\mathsf{pre}}$ and $[\![\cdot]\!]_{\widehat{\mathsf{pre}}}$, that have analogous proof obligations. We also prove that they correctly interpret nonrecursive, nonprobabilistic programs.

In the full development [28], two more semantic functions interpret programs as extensional functions, which are used to prove measurability. Another semantic function collects information needed for advanced Monte Carlo algorithms. In all, we have 9 related semantic functions, each defined by 11 or 12 rules, whose correctness and relationships must be proved by structural induction. Doing so is tedious and error-prone. We need a way to parameterize one semantic function on many meanings, where each "meaning" is simpler than a semantic function and ideally has exploitable properties.

Moggi [22] introduced monads as a categorical "metalanguage" for interpreting programs. Wadler [31] showed how to use monad categories in pure functional programming to encode and hide side effects such as mutation and randomness. Haskell programmers now primarily encode programs with side effects using **do-notation**, which is transformed into any monad. Essentially, Haskell has a built-in semantic function parameterized on a monad.

Other researchers have identified arrows [10] and idioms [19] as useful kinds of categories. Different kinds of categories are good for encoding different kinds of effects, and have different levels of expressiveness [16]. Arrows are good categories for interpreting first-order languages. We therefore interpret programs 9 different ways by parameterizing a semantic function on one of 9 arrow categories.

In our formulation, an arrow category consists of a type constructor and five combinators; each is thus half as complicated as the semantic function. Their categorical properties also allow two drastic simplifications. First, they allow proving the correctness of a semantic function $[\![\cdot]\!]_{\mathsf{b}}$ with respect to $[\![\cdot]\!]_{\mathsf{a}}$ by proving a simple theorem about arrows $\mathsf{a}$ and $\mathsf{b}$. Second, they allow us to *derive* all the arrows for recursive, probabilistic programs at once, by lifting the arrows for nonrecursive, nonprobabilistic programs.

### 2.5 Types and Notation

Because some arrows carry out uncountably infinite computations, we must define their combinators in a sufficiently powerful $\lambda$-calculus. We use $\lambda_{\mathrm{ZFC}}$ [29].

Though $\lambda_{\mathrm{ZFC}}$ is untyped, it helps to use a manually checked, auxiliary type system. For example, the types of some of $\lambda_{\mathrm{ZFC}}$'s primitives are those of membership $(\in) : \mathsf{x} \to \mathsf{Set}\ \mathsf{x} \to \mathsf{Bool}$, powerset $\mathcal{P} : \mathsf{Set}\ \mathsf{x} \to \mathsf{Set}\ (\mathsf{Set}\ \mathsf{x})$, big union $\bigcup : \mathsf{Set}\ (\mathsf{Set}\ \mathsf{x}) \to \mathsf{Set}\ \mathsf{x}$, and the $\mathtt{map}$-like $\mathsf{image} : (\mathsf{x} \to \mathsf{y}) \to \mathsf{Set}\ \mathsf{x} \to \mathsf{Set}\ \mathsf{y}$. We allow sets to be used as types, as in $\mathsf{max} : \langle \mathbb{R}, \mathbb{R} \rangle \to \mathbb{R}$.

More precisely, types are characterized by these rules:

- $\mathsf{x} \to \mathsf{y}$ is the type of intensional, partial functions from type $\mathsf{x}$ to type $\mathsf{y}$.
- $\langle \mathsf{x}, \mathsf{y} \rangle$ is the type of pairs of values with types $\mathsf{x}$ and $\mathsf{y}$.
- $\mathsf{Set}\ \mathsf{x}$ is the type of sets whose members have type $\mathsf{x}$.

– An uppercase type variable such as $\mathsf{X}$ represents a set used as a type.

Because the inhabitants of the type $\mathsf{Set}\ \mathsf{X}$ and $\mathcal{P}\ \mathsf{X}$ (i.e. subsets of the set $\mathsf{X}$) are the same, they are equivalent types. Similarly, $\langle \mathsf{X}, \mathsf{Y} \rangle$ is equivalent to $\mathsf{X} \times \mathsf{Y}$.

Type constructors are defined using '::='; e.g. $\mathsf{X} \rightsquigarrow_{\perp} \mathsf{Y} ::= \mathsf{X} \to (\mathsf{Y} \cup \{\perp\})$.

The set $\mathsf{X}^{\mathsf{J}}$ contains all extensional, total functions from set $\mathsf{J}$ to set $\mathsf{X}$; i.e. vectors of $\mathsf{X}$ indexed by $\mathsf{J}$. We use adjacency (i.e. $\mathsf{f}\ \mathsf{a}$) to apply both intensional and extensional functions. For example, the first element of $\mathsf{f} : [0,1]^{\mathbb{N}}$ is $\mathsf{f}\ 0$.

Proofs, which we elide to save space, are in the first author's dissertation [28].

## 3 Arrows and First-Order Semantics

This section presents the categorical tools we use to derive many semantics from a single standard semantics in a way that makes them easy to prove correct.

Arrows [10], like monads [31], thread effects through computations in a way that imposes structure. But arrow computations are always

– Function-like. The type constructor for arrow $\mathsf{a}$ is written $\mathsf{x} \rightsquigarrow_{\mathsf{a}} \mathsf{y}$ to connote this. In fact, the *function arrow*'s type constructor is $\mathsf{x} \rightsquigarrow \mathsf{y} ::= \mathsf{x} \to \mathsf{y}$.
– First-order. There is no way to derive the higher-order application combinator $\mathsf{app} : \langle \mathsf{x} \rightsquigarrow_{\mathsf{a}} \mathsf{y}, \mathsf{x} \rangle \rightsquigarrow_{\mathsf{a}} \mathsf{y}$ from the combinators that define arrow $\mathsf{a}$.

The first property makes arrows a good fit for a compositional translation from expressions to pure functions that operate on random sources. The second property makes arrows a good fit for the semantics of a first-order language.

### 3.1 Arrow Combinators and Laws

Arrows factor computation into the following tasks: (1) referring to pure, primitive functions, (2) applying primitive or first-order functions, (3) binding values to local variables and creating data structures, and (4) branching based on the results of prior computations. The first four arrow combinators correspond respectively with each of these tasks. A fifth combinator allows lazy branching in a call-by-value language such as $\lambda_{\text{ZFC}}$.

For laziness, we need a singleton type for thunks. We use the set $1 := \{0\}$.

**Definition 1 (arrow[3]).** *A binary type constructor* $(\rightsquigarrow_{\mathsf{a}})$ *and the combinators*

$$
\begin{aligned}
\mathsf{arr_a} &: (\mathsf{x} \to \mathsf{y}) \to (\mathsf{x} \rightsquigarrow_{\mathsf{a}} \mathsf{y}) && \text{lift} \\
(\ggg_{\mathsf{a}}) &: (\mathsf{x} \rightsquigarrow_{\mathsf{a}} \mathsf{y}) \to (\mathsf{y} \rightsquigarrow_{\mathsf{a}} \mathsf{z}) \to (\mathsf{x} \rightsquigarrow_{\mathsf{a}} \mathsf{z}) && \text{compose} \\
(\&\&\&_{\mathsf{a}}) &: (\mathsf{x} \rightsquigarrow_{\mathsf{a}} \mathsf{y}) \to (\mathsf{x} \rightsquigarrow_{\mathsf{a}} \mathsf{z}) \to (\mathsf{x} \rightsquigarrow_{\mathsf{a}} \langle \mathsf{y}, \mathsf{z} \rangle) && \text{pair} \\
\mathsf{ifte_a} &: (\mathsf{x} \rightsquigarrow_{\mathsf{a}} \mathsf{Bool}) \to (\mathsf{x} \rightsquigarrow_{\mathsf{a}} \mathsf{y}) \to (\mathsf{x} \rightsquigarrow_{\mathsf{a}} \mathsf{y}) \to (\mathsf{x} \rightsquigarrow_{\mathsf{a}} \mathsf{y}) && \text{if-then-else} \\
\mathsf{lazy_a} &: (1 \to (\mathsf{x} \rightsquigarrow_{\mathsf{a}} \mathsf{y})) \to (\mathsf{x} \rightsquigarrow_{\mathsf{a}} \mathsf{y}) && \text{laziness}
\end{aligned}
$$

*define an* ***arrow*** *if certain monoid, homomorphism, and other laws hold [10].*

[3] These are actually arrows *with choice*, which are typically defined using $\mathsf{first_a}$ and $\mathsf{left_a}$ instead of $(\&\&\&_{\mathsf{a}})$ and $\mathsf{ifte_a}$. We find $\mathsf{ifte_a}$ more natural for semantics than $\mathsf{left_a}$, and $(\&\&\&_{\mathsf{a}})$ better matches the pairing preimage identity in (12).

For example, the **function arrow** is defined by the type constructor $\mathsf{x} \rightsquigarrow \mathsf{y} ::= \mathsf{x} \to \mathsf{y}$ and the combinators

$$\begin{aligned} \mathsf{arr}\ \mathsf{f} &:= \mathsf{f} \\ (\mathsf{f}_1 \ggg \mathsf{f}_2)\ \mathsf{a} &:= \mathsf{f}_2\ (\mathsf{f}_1\ \mathsf{a}) \\ (\mathsf{f}_1\ \&\&\&\ \mathsf{f}_2)\ \mathsf{a} &:= \langle \mathsf{f}_1\ \mathsf{a}, \mathsf{f}_2\ \mathsf{a} \rangle \\ \mathsf{ifte}\ \mathsf{f}_1\ \mathsf{f}_2\ \mathsf{f}_3\ \mathsf{a} &:= \mathsf{if}\ \mathsf{f}_1\ \mathsf{a}\ \mathsf{then}\ \mathsf{f}_2\ \mathsf{a}\ \mathsf{else}\ \mathsf{f}_3\ \mathsf{a} \\ \mathsf{lazy}\ \mathsf{f}\ \mathsf{a} &:= \mathsf{f}\ 0\ \mathsf{a} \end{aligned} \tag{13}$$

To demonstrate compositionally interpreting probabilistic programs as arrow computations, we interpret $\mathsf{max}\ \langle 0.5, \mathsf{random} \rangle$ as a function arrow computation $\mathsf{f} : [0,1] \rightsquigarrow \mathbb{R}$. For any random source $\mathsf{r} \in [0,1]$, the interpretation of 0.5 should return 0.5, so 0.5 means $\lambda \mathsf{r}.\, 0.5$, or $\mathsf{const}\ 0.5$ where $\mathsf{const}\ \mathsf{v} := \lambda_.\, \mathsf{v}$. Assuming $\mathsf{r} \in [0,1]$ is uniformly distributed, $\mathsf{random}$ means $\lambda \mathsf{r}.\, \mathsf{r}$, or $\mathsf{id}$. We use $(\&\&\&)$ to apply each of these interpretations to the random source to create a pair, and $(\ggg)$ to send the pair to $\mathsf{max}$. Thus, $\mathsf{max}\ \langle 0.5, \mathsf{random} \rangle$, interpreted as a function arrow computation, is $\mathsf{f} := ((\mathsf{const}\ 0.5)\ \&\&\&\ \mathsf{id}) \ggg \mathsf{max}$.

By substituting the definitions of $\mathsf{const}$, $\mathsf{id}$, $(\&\&\&)$ and $(\ggg)$, we would find that $\mathsf{f}$ is equivalent to $\lambda \mathsf{r}.\, \mathsf{max}\ \langle 0.5, \mathsf{r} \rangle$, similar to the interpretation in (8).

Only the function arrow can so cavalierly use pure functions as arrow computations. In any other arrow $\mathsf{a}$, pure functions must be *lifted* using $\mathsf{arr}_\mathsf{a}$, to allow the arrow to manage any state or effects. Therefore, the interpretation of $\mathsf{max}\ \langle 0.5, \mathsf{random} \rangle$ as an arrow $\mathsf{a}$ computation $\mathsf{f}_\mathsf{a} : [0,1] \rightsquigarrow_\mathsf{a} \mathbb{R}$ is

$$\mathsf{f}_\mathsf{a} := (\mathsf{arr}_\mathsf{a}\ (\mathsf{const}\ 0.5)\ \&\&\&_\mathsf{a}\ \mathsf{arr}_\mathsf{a}\ \mathsf{id}) \ggg_\mathsf{a} \mathsf{arr}_\mathsf{a}\ \mathsf{max} \tag{14}$$

So far, we have ignored the many arrow laws, which ensure that arrows are well-behaved (e.g. effects are correctly ordered) and are useful in proofs of theorems that quantify over arrows (i.e. nothing else is known about them). Fortunately, we can prove all the laws for an arrow $\mathsf{b}$ by defining it in terms of an arrow $\mathsf{a}$ for which the laws hold, and proving two properties about the lift from $\mathsf{a}$ to $\mathsf{b}$. The first property is that the lift from $\mathsf{a}$ to $\mathsf{b}$ is distributive.

**Definition 2 (arrow homomorphism).** $\mathsf{lift}_\mathsf{b} : (\mathsf{x} \rightsquigarrow_\mathsf{a} \mathsf{y}) \to (\mathsf{x} \rightsquigarrow_\mathsf{b} \mathsf{y})$ *is an* ***arrow homomorphism*** *from* $\mathsf{a}$ *to* $\mathsf{b}$ *if these distributive laws hold:*

$$\mathsf{lift}_\mathsf{b}\ (\mathsf{arr}_\mathsf{a}\ \mathsf{f}) \equiv \mathsf{arr}_\mathsf{b}\ \mathsf{f} \tag{15}$$

$$\mathsf{lift}_\mathsf{b}\ (\mathsf{f}_1 \ggg_\mathsf{a} \mathsf{f}_2) \equiv (\mathsf{lift}_\mathsf{b}\ \mathsf{f}_1) \ggg_\mathsf{b} (\mathsf{lift}_\mathsf{b}\ \mathsf{f}_2) \tag{16}$$

$$\mathsf{lift}_\mathsf{b}\ (\mathsf{f}_1\ \&\&\&_\mathsf{a}\ \mathsf{f}_2) \equiv (\mathsf{lift}_\mathsf{b}\ \mathsf{f}_1)\ \&\&\&_\mathsf{b}\ (\mathsf{lift}_\mathsf{b}\ \mathsf{f}_2) \tag{17}$$

$$\mathsf{lift}_\mathsf{b}\ (\mathsf{ifte}_\mathsf{a}\ \mathsf{f}_1\ \mathsf{f}_2\ \mathsf{f}_3) \equiv \mathsf{ifte}_\mathsf{b}\ (\mathsf{lift}_\mathsf{b}\ \mathsf{f}_1)\ (\mathsf{lift}_\mathsf{b}\ \mathsf{f}_2)\ (\mathsf{lift}_\mathsf{b}\ \mathsf{f}_3) \tag{18}$$

$$\mathsf{lift}_\mathsf{b}\ (\mathsf{lazy}_\mathsf{a}\ \mathsf{f}) \equiv \mathsf{lazy}_\mathsf{b}\ \lambda 0.\, \mathsf{lift}_\mathsf{b}\ (\mathsf{f}\ 0) \tag{19}$$

*where "$\equiv$" is an arrow-specific equivalence relation.*

The second property is that the lift is right-invertible (i.e. surjective).

**Theorem 1 (right-invertible homomorphism implies arrow laws).** *If* $\mathsf{lift}_\mathsf{b} : (\mathsf{x} \rightsquigarrow_\mathsf{a} \mathsf{y}) \to (\mathsf{x} \rightsquigarrow_\mathsf{b} \mathsf{y})$ *is a right-invertible homomorphism from* $\mathsf{a}$ *to* $\mathsf{b}$ *and the arrow laws hold for* $\mathsf{a}$*, then the arrow laws hold for* $\mathsf{b}$*.*

$$
\begin{aligned}
p &::\equiv f := e;\ \dots\ ; e \\
e &::\equiv \mathsf{let}\ e\ e \mid \mathsf{env}\ n \mid \mathsf{if}\ e\ \mathsf{then}\ e\ \mathsf{else}\ e \mid \langle e, e\rangle \mid f\ e \mid \delta\ e \mid v \\
f &::\equiv \textit{(first-order function names)} \\
\delta &::\equiv \textit{(primitive function names)} \\
v &::\equiv \langle v, v\rangle \mid \langle\rangle \mid \mathsf{true} \mid \mathsf{false} \mid \textit{(other first-order constants)}
\end{aligned}
$$

$$
\begin{aligned}
[\![ f := e;\ \dots\ ; e_b ]\!]_{\mathsf{a}} &:\equiv f := [\![ e ]\!]_{\mathsf{a}};\ \dots\ ; [\![ e_b ]\!]_{\mathsf{a}} \\
[\![ \mathsf{let}\ e\ e_b ]\!]_{\mathsf{a}} &:\equiv ([\![ e ]\!]_{\mathsf{a}}\ \&\&\&_{\mathsf{a}}\ \mathsf{arr_a}\ \mathsf{id}) \ggg_{\mathsf{a}} [\![ e_b ]\!]_{\mathsf{a}} \\
[\![ \mathsf{env}\ 0 ]\!]_{\mathsf{a}} &:\equiv \mathsf{arr_a}\ \mathsf{fst} \\
[\![ \mathsf{env}\ (n+1) ]\!]_{\mathsf{a}} &:\equiv \mathsf{arr_a}\ \mathsf{snd} \ggg_{\mathsf{a}} [\![ \mathsf{env}\ n ]\!]_{\mathsf{a}} \\
[\![ \mathsf{if}\ e_c\ \mathsf{then}\ e_t\ \mathsf{else}\ e_f ]\!]_{\mathsf{a}} &:\equiv \mathsf{ifte_a}\ [\![ e_c ]\!]_{\mathsf{a}}\ (\mathsf{lazy_a}\ \lambda 0.\, [\![ e_t ]\!]_{\mathsf{a}})\ (\mathsf{lazy_a}\ \lambda 0.\, [\![ e_f ]\!]_{\mathsf{a}})
\end{aligned}
\qquad
\begin{aligned}
[\![ \langle e_1, e_2\rangle ]\!]_{\mathsf{a}} &:\equiv [\![ e_1 ]\!]_{\mathsf{a}}\ \&\&\&_{\mathsf{a}}\ [\![ e_2 ]\!]_{\mathsf{a}} \\
[\![ f\ e ]\!]_{\mathsf{a}} &:\equiv [\![ \langle e, \langle\rangle\rangle ]\!]_{\mathsf{a}} \ggg_{\mathsf{a}} f \\
[\![ \delta\ e ]\!]_{\mathsf{a}} &:\equiv [\![ e ]\!]_{\mathsf{a}} \ggg_{\mathsf{a}} \mathsf{arr_a}\ \delta \\
[\![ v ]\!]_{\mathsf{a}} &:\equiv \mathsf{arr_a}\ (\mathsf{const}\ v)
\end{aligned}
$$

$$
\text{where}\quad \begin{aligned} \mathsf{const}\ \mathsf{v} &:= \lambda_.\,\mathsf{v} \\ \mathsf{id} &:= \lambda \mathsf{v}.\,\mathsf{v} \end{aligned} \qquad \text{subject to}\quad [\![ p ]\!]_{\mathsf{a}} : \langle\rangle \rightsquigarrow_{\mathsf{a}} \mathsf{y} \text{ for some } \mathsf{y}
$$

Fig. 2: Interpretation of a let-calculus with first-order definitions and De-Bruijn-indexed bindings as arrow a computations. Here, '$::\equiv$' denotes definitional extension for grammars and '$:\equiv$' denotes definitional extension for syntax.

### 3.2 First-Order Let-Calculus Semantics

Figure 2 defines a semantic function $[\![\cdot]\!]_{\mathsf{a}}$ that interprets first-order programs as arrow computations for any arrow a. A program is a sequence of function definitions separated by semicolons (or line breaks), followed by a final expression. Function definitions may be mutually recursive because they are interpreted as definitions in a metalanguage in which mutual recursion is supported. (We thus do not need an explicit fixpoint operator.) Unlike functions, local variables are unnamed: we use De Bruijn indexes, with 0 referring to the innermost binding.

The result of applying $[\![\cdot]\!]_{\mathsf{a}}$ is a $\lambda_{\mathrm{ZFC}}$ program in **environment-passing style** where the environment is a stack. The final expression has type $\langle\rangle \rightsquigarrow_{\mathsf{a}} \mathsf{y}$, where y is the type of the program's output and $\langle\rangle$ denotes the empty stack. A let expression uses pairing ($\&\&\&_{\mathsf{a}}$) to push a value onto the stack and composition ($\ggg_{\mathsf{a}}$) to pass the resulting stack to its body. First-order functions have type $\langle \mathsf{x}, \langle\rangle\rangle \rightsquigarrow_{\mathsf{a}} \mathsf{y}$ where x is the argument type and y is the return type. Application passes a stack containing just an x using pairing and composition.

Using De Bruijn indexes, $\mathsf{g}\ \mathsf{x} := \mathsf{g}\ \mathsf{x}$ is written $\mathsf{g} := \mathsf{g}\ (\mathsf{env}\ 0)$, which $[\![\cdot]\!]_{\mathsf{a}}$ interprets as $\mathsf{g} := [\![\langle \mathsf{env}\ 0, \langle\rangle\rangle]\!]_{\mathsf{a}} \ggg_{\mathsf{a}} \mathsf{g}$. To disallow such circular definitions, and ill-typed expressions like $\mathsf{max}\ \langle 0.5, \langle\rangle\rangle$, we require programs to be **well-defined**.

**Definition 3 (well-defined).** *An expression (or program) $e$ is **well-defined** under arrow* a *if $[\![e]\!]_{\mathsf{a}}$ terminates and $[\![e]\!]_{\mathsf{a}} : \mathsf{x} \rightsquigarrow_{\mathsf{a}} \mathsf{y}$ for some* x *and* y.

Well-definedness guarantees that recursion is guarded by if expressions, as $[\![\mathsf{if}\ e_c\ \mathsf{then}\ e_t\ \mathsf{else}\ e_f]\!]_{\mathsf{a}}$ wraps $[\![e_t]\!]_{\mathsf{a}}$ and $[\![e_f]\!]_{\mathsf{a}}$ in thunks. It does *not* guarantee that *running* an interpretation always terminates. For example, the program $\mathsf{g} := \mathsf{if}\ \mathsf{true}\ \mathsf{then}\ \mathsf{g}\ (\mathsf{env}\ 0)\ \mathsf{else}\ 0;\ \mathsf{g}\ 0$ is well-defined under the function arrow, but applying its interpretation to $\langle\rangle$ does not terminate. Section 5 deals with such programs by defining arrows that take finitely many branches, or return $\bot$.

$$
\begin{array}{ll}
\mathsf{X} \rightsquigarrow_{\bot} \mathsf{Y} ::= \mathsf{X} \to \mathsf{Y}_{\bot} & \mathsf{ifte}_{\bot}\ \mathsf{f}_1\ \mathsf{f}_2\ \mathsf{f}_3\ \mathsf{a} := \mathsf{case}\ \mathsf{f}_1\ \mathsf{a}\ \mathsf{of} \\
 & \qquad\qquad\qquad\quad \mathsf{true} \longrightarrow \mathsf{f}_2\ \mathsf{a} \\
\mathsf{arr}_{\bot}\ \mathsf{f}\ \mathsf{a} := \mathsf{f}\ \mathsf{a} & \qquad\qquad\qquad\quad \mathsf{false} \longrightarrow \mathsf{f}_3\ \mathsf{a} \\
(\mathsf{f}_1 \ggg_{\bot} \mathsf{f}_2)\ \mathsf{a} := \mathsf{case}\ \mathsf{f}_1\ \mathsf{a}\ \mathsf{of} & \qquad\qquad\qquad\quad \bot \longrightarrow \bot \\
\qquad\qquad\qquad \bot \longrightarrow \bot & (\mathsf{f}_1\ \&\&\&_{\bot}\ \mathsf{f}_2)\ \mathsf{a} := \mathsf{case}\ \langle \mathsf{f}_1\ \mathsf{a}, \mathsf{f}_2\ \mathsf{a} \rangle\ \mathsf{of} \\
\qquad\qquad\qquad \mathsf{b} \longrightarrow \mathsf{f}_2\ \mathsf{b} & \qquad\qquad\qquad \langle \bot, _ \rangle \longrightarrow \bot \\
 & \qquad\qquad\qquad \langle _, \bot \rangle \longrightarrow \bot \\
\mathsf{lazy}_{\bot}\ \mathsf{f}\ \mathsf{a} := \mathsf{f}\ 0\ \mathsf{a} & \qquad\qquad\qquad \langle \mathsf{b}_1, \mathsf{b}_2 \rangle \longrightarrow \langle \mathsf{b}_1, \mathsf{b}_2 \rangle
\end{array}
$$

Fig. 3: Bottom arrow definitions.

Most of our semantic correctness results rely on the following theorem.

**Theorem 2 (homomorphisms distribute over expressions).** *Let* $\mathsf{lift_b} : (\mathsf{x} \rightsquigarrow_{\mathsf{a}} \mathsf{y}) \to (\mathsf{x} \rightsquigarrow_{\mathsf{b}} \mathsf{y})$ *be an arrow homomorphism. For all* $e$*,* $[\![e]\!]_{\mathsf{b}} \equiv \mathsf{lift_b}\ [\![e]\!]_{\mathsf{a}}$.

Much of our development proceeds in the following way.

1. Define an arrow $\mathsf{a}$ to interpret programs using $[\![\cdot]\!]_{\mathsf{a}}$.
2. Define $\mathsf{lift_b} : (\mathsf{x} \rightsquigarrow_{\mathsf{a}} \mathsf{y}) \to (\mathsf{x} \rightsquigarrow_{\mathsf{b}} \mathsf{y})$ from arrow $\mathsf{a}$ to $\mathsf{b}$ with the property that if $\mathsf{f} : \mathsf{x} \rightsquigarrow_{\mathsf{a}} \mathsf{y}$, then $\mathsf{lift_b}\ \mathsf{f}$ is correct.
3. Prove $\mathsf{lift_b}$ is a homomorphism; therefore $[\![e]\!]_{\mathsf{b}}$ is correct (Theorem 2).
4. Prove $\mathsf{lift_b}$ is right-invertible; therefore $\mathsf{b}$ obeys the arrow laws (Theorem 1).

In shorter terms, *if* $\mathsf{b}$ *is defined in terms of a right-invertible homomorphism from arrow* $\mathsf{a}$ *to* $\mathsf{b}$*, then* $[\![\cdot]\!]_{\mathsf{b}}$ *is correct with respect to* $[\![\cdot]\!]_{\mathsf{a}}$.

# 4 The Bottom and Preimage Arrows

The following commutative diagram shows the relationships between the arrows $\mathsf{X} \rightsquigarrow_{\bot} \mathsf{Y}$ and $\mathsf{X} \rightsquigarrow_{\mathsf{pre}} \mathsf{Y}$ for interpreting nonrecursive, nonprobabilistic programs, and $\mathsf{X} \rightsquigarrow_{\bot^*} \mathsf{Y}$ and $\mathsf{X} \rightsquigarrow_{\mathsf{pre}^*} \mathsf{Y}$ for interpreting recursive, probabilistic programs.

$$
\begin{array}{ccc}
\mathsf{X} \rightsquigarrow_{\bot} \mathsf{Y} & \xrightarrow{\mathsf{lift_{pre}}} & \mathsf{X} \rightsquigarrow_{\mathsf{pre}} \mathsf{Y} \\
\eta_{\bot^*} \downarrow & & \downarrow \eta_{\mathsf{pre}^*} \\
\mathsf{X} \rightsquigarrow_{\bot^*} \mathsf{Y} & \xrightarrow[\mathsf{lift_{pre^*}}]{} & \mathsf{X} \rightsquigarrow_{\mathsf{pre}^*} \mathsf{Y}
\end{array}
\tag{20}
$$

In this section, we define the top row.

## 4.1 The Bottom Arrow

To use Theorem 2 to prove correct the interpretations of expressions as preimage arrow computations, we need to define the preimage arrow in terms of a simpler arrow with easily understood behavior. The function arrow (13) is an obvious candidate. However, we will need to represent possible nontermination as an error value, so we need a slightly more complicated arrow.

Fig. 3 defines the **bottom arrow**, which is similar to the function arrow but propagates the error value $\bot$. Its computations have type $\mathsf{X} \rightsquigarrow_{\bot} \mathsf{Y} ::= \mathsf{X} \to \mathsf{Y}_{\bot}$, where $\mathsf{Y}_{\bot} ::= \mathsf{Y} \cup \{\bot\}$.

To prove the arrow laws, we need coarse enough notion of equivalence.

$$\mathsf{X} \rightharpoonup_{\mathsf{pre}} \mathsf{Y} ::= \langle \mathsf{Set\ Y}, \mathsf{Set\ Y} \to \mathsf{Set\ X} \rangle$$

$$\begin{aligned}
&\mathsf{pre} : (\mathsf{X} \rightsquigarrow_\bot \mathsf{Y}) \to \mathsf{Set\ X} \to (\mathsf{X} \rightharpoonup_{\mathsf{pre}} \mathsf{Y}) \\
&\mathsf{pre\ f\ A} \ := \ \langle \mathsf{image}_\bot\ \mathsf{f\ A}, \mathsf{preimage}_\bot\ \mathsf{f\ A} \rangle
\end{aligned}$$

$$\varnothing_{\mathsf{pre}} \ := \ \langle \varnothing, \lambda \mathsf{B}.\, \varnothing \rangle$$

$$\begin{aligned}
&\mathsf{ap}_{\mathsf{pre}} : (\mathsf{X} \rightharpoonup_{\mathsf{pre}} \mathsf{Y}) \to \mathsf{Set\ Y} \to \mathsf{Set\ X} \\
&\mathsf{ap}_{\mathsf{pre}}\ \langle \mathsf{B}', \mathsf{p} \rangle\ \mathsf{B} \ := \ \mathsf{p}\ (\mathsf{B} \cap \mathsf{B}')
\end{aligned}$$

$$\begin{aligned}
&\mathsf{range}_{\mathsf{pre}} : (\mathsf{X} \rightharpoonup_{\mathsf{pre}} \mathsf{Y}) \to \mathsf{Set\ Y} \\
&\mathsf{range}_{\mathsf{pre}}\ \langle \mathsf{B}', \mathsf{p} \rangle \ := \ \mathsf{B}'
\end{aligned}$$

$$\begin{aligned}
&\langle \cdot, \cdot \rangle_{\mathsf{pre}} : (\mathsf{X} \rightharpoonup_{\mathsf{pre}} \mathsf{Y}_1) \to (\mathsf{X} \rightharpoonup_{\mathsf{pre}} \mathsf{Y}_2) \to (\mathsf{X} \rightharpoonup_{\mathsf{pre}} \langle \mathsf{Y}_1, \mathsf{Y}_2 \rangle) \\
&\langle \langle \mathsf{B}'_1, \mathsf{p}_1 \rangle, \langle \mathsf{B}'_2, \mathsf{p}_2 \rangle \rangle_{\mathsf{pre}} \ := \\
&\quad \mathsf{let}\ \ \mathsf{B}' := \mathsf{B}'_1 \times \mathsf{B}'_2 \\
&\qquad\quad \mathsf{p} := \lambda \mathsf{B}. \bigcup_{\langle \mathsf{b}_1, \mathsf{b}_2 \rangle \in \mathsf{B}} (\mathsf{p}_1\ \{\mathsf{b}_1\}) \cap (\mathsf{p}_2\ \{\mathsf{b}_2\}) \\
&\quad \mathsf{in}\ \ \langle \mathsf{B}', \mathsf{p} \rangle
\end{aligned}$$

$$\begin{aligned}
&(\circ_{\mathsf{pre}}) : (\mathsf{Y} \rightharpoonup_{\mathsf{pre}} \mathsf{Z}) \to (\mathsf{X} \rightharpoonup_{\mathsf{pre}} \mathsf{Y}) \to (\mathsf{X} \rightharpoonup_{\mathsf{pre}} \mathsf{Z}) \\
&\langle \mathsf{C}', \mathsf{p}_2 \rangle \circ_{\mathsf{pre}} \mathsf{h}_1 \ := \ \langle \mathsf{C}', \lambda \mathsf{C}.\, \mathsf{ap}_{\mathsf{pre}}\ \mathsf{h}_1\ (\mathsf{p}_2\ \mathsf{C}) \rangle
\end{aligned}$$

$$\begin{aligned}
&(\cup_{\mathsf{pre}}) : (\mathsf{X} \rightharpoonup_{\mathsf{pre}} \mathsf{Y}) \to (\mathsf{X} \rightharpoonup_{\mathsf{pre}} \mathsf{Y}) \to (\mathsf{X} \rightharpoonup_{\mathsf{pre}} \mathsf{Y}) \\
&\langle \mathsf{B}'_1, \mathsf{p}_1 \rangle \cup_{\mathsf{pre}} \langle \mathsf{B}'_2, \mathsf{p}_2 \rangle \ := \\
&\quad \langle \mathsf{B}'_1 \cup \mathsf{B}'_2, \lambda \mathsf{B}.\, \mathsf{ap}_{\mathsf{pre}}\ \langle \mathsf{B}'_1, \mathsf{p}_1 \rangle\ \mathsf{B} \cup \mathsf{ap}_{\mathsf{pre}}\ \langle \mathsf{B}'_2, \mathsf{p}_2 \rangle\ \mathsf{B} \rangle
\end{aligned}$$

$$\begin{aligned}
&\mathsf{image}_\bot : (\mathsf{X} \rightsquigarrow_\bot \mathsf{Y}) \to \mathsf{Set\ X} \to \mathsf{Set\ Y} \\
&\mathsf{image}_\bot\ \mathsf{f\ A} \ := \ (\mathsf{image\ f\ A}) \backslash \{\bot\}
\end{aligned}$$

$$\begin{aligned}
&\mathsf{preimage}_\bot : (\mathsf{X} \rightsquigarrow_\bot \mathsf{Y}) \to \mathsf{Set\ X} \to \mathsf{Set\ Y} \to \mathsf{Set\ X} \\
&\mathsf{preimage}_\bot\ \mathsf{f\ A\ B} \ := \ \{\mathsf{a} \in \mathsf{A} \mid \mathsf{f\ a} \in \mathsf{B}\}
\end{aligned}$$

Fig. 4: Preimage functions and operations.

**Definition 4 (bottom arrow equivalence).** *Two computations* $\mathsf{f}_1 : \mathsf{X} \rightsquigarrow_\bot \mathsf{Y}$ *and* $\mathsf{f}_2 : \mathsf{X} \rightsquigarrow_\bot \mathsf{Y}$ *are equivalent, or* $\mathsf{f}_1 \equiv \mathsf{f}_2$*, when* $\mathsf{f}_1\ \mathsf{a} \equiv \mathsf{f}_2\ \mathsf{a}$ *for all* $\mathsf{a} \in \mathsf{X}$.

Using bottom arrow equivalence, it is easy to show that $(\rightsquigarrow_\bot)$ is isomorphic to the Maybe monad's Kleisli arrow. By Theorem 1, the arrow laws hold.

### 4.2 The Preimage Function Type and Operations

Before defining the preimage arrow, we need a type of preimage functions. $\mathsf{Set\ Y} \to \mathsf{Set\ X}$ would be a good candidate, except that the $(\ggg_{\mathsf{pre}})$ combinator will require preimage functions to have observable domains, but instances of $\mathsf{Set\ Y} \to \mathsf{Set\ X}$ are intensional functions. We therefore define

$$\mathsf{X} \rightharpoonup_{\mathsf{pre}} \mathsf{Y} \ ::= \ \langle \mathsf{Set\ Y}, \mathsf{Set\ Y} \to \mathsf{Set\ X} \rangle \tag{21}$$

as the type of preimage functions. Fig. 4 defines the necessary operations on them. Operations $\langle \cdot, \cdot \rangle_{\mathsf{pre}}$ and $(\circ_{\mathsf{pre}})$ return preimage functions that compute preimages under pairing and composition, and are derived from the preimage identities in (12); $(\cup_{\mathsf{pre}})$ computes unions and is used to define $\mathsf{ifte}_{\mathsf{pre}}$.

Fig. 4 also defines $\mathsf{image}_\bot$ and $\mathsf{preimage}_\bot$ to operate on bottom arrow computations: $\mathsf{image}_\bot\ \mathsf{f\ A}$ computes $\mathsf{f}$'s range (with domain $\mathsf{A}$), and $\mathsf{preimage}_\bot\ \mathsf{f\ A}$ returns a function that computes preimages under $\mathsf{f}$ restricted to $\mathsf{A}$. Together, they can be used to convert bottom arrow computations to preimage functions:

$$\begin{aligned}
&\mathsf{pre} : (\mathsf{X} \rightsquigarrow_\bot \mathsf{Y}) \to \mathsf{Set\ X} \to (\mathsf{X} \rightharpoonup_{\mathsf{pre}} \mathsf{Y}) \\
&\mathsf{pre\ f\ A} \ := \ \langle \mathsf{image}_\bot\ \mathsf{f\ A}, \mathsf{preimage}_\bot\ \mathsf{f\ A} \rangle
\end{aligned} \tag{22}$$

Lastly, the $\mathsf{ap}_{\mathsf{pre}}$ function in Fig. 4 applies a preimage function to a set.

Preimage arrow correctness depends on $\mathsf{ap}_{\mathsf{pre}}$ and $\mathsf{pre}$ behaving like $\mathsf{preimage}_\bot$.

$$
\begin{aligned}
\mathsf{X} \rightsquigarrow_{\mathsf{pre}} \mathsf{Y} &::= \mathsf{Set\ X} \to (\mathsf{X} \rightharpoonup_{\mathsf{pre}} \mathsf{Y}) \\
\mathsf{arr_{pre}} &:= \mathsf{lift_{pre}} \circ \mathsf{arr_{\perp}} \\
(\mathsf{h_1} \ggg_{\mathsf{pre}} \mathsf{h_2})\ \mathsf{A} &:= \mathsf{let}\ \ \mathsf{h_1'} := \mathsf{h_1\ A} \\
&\qquad\quad \mathsf{h_2'} := \mathsf{h_2}\ (\mathsf{range_{pre}}\ \mathsf{h_1'}) \\
&\quad\ \ \mathsf{in}\ \ \mathsf{h_2'} \circ_{\mathsf{pre}} \mathsf{h_1'} \\
(\mathsf{h_1}\ \&\&\&_{\mathsf{pre}}\ \mathsf{h_2})\ \mathsf{A} &:= \langle \mathsf{h_1\ A}, \mathsf{h_2\ A} \rangle_{\mathsf{pre}}
\end{aligned}
$$

$$
\begin{aligned}
&\mathsf{ifte_{pre}}\ \mathsf{h_1\ h_2\ h_3\ A} := \\
&\quad \mathsf{let}\ \ \mathsf{h_1'} := \mathsf{h_1\ A} \\
&\qquad\quad \mathsf{h_2'} := \mathsf{h_2}\ (\mathsf{ap_{pre}}\ \mathsf{h_1'}\ \{\mathsf{true}\}) \\
&\qquad\quad \mathsf{h_3'} := \mathsf{h_3}\ (\mathsf{ap_{pre}}\ \mathsf{h_1'}\ \{\mathsf{false}\}) \\
&\quad \mathsf{in}\ \ \mathsf{h_2'} \cup_{\mathsf{pre}} \mathsf{h_3'} \\
&\mathsf{lazy_{pre}}\ \mathsf{h\ A} := \mathsf{if\ A} = \varnothing\ \mathsf{then}\ \varnothing_{\mathsf{pre}}\ \mathsf{else\ h\ 0\ A} \\
&\mathsf{lift_{pre}} := \mathsf{pre}
\end{aligned}
$$

Fig. 5: Preimage arrow definitions.

**Theorem 3** ($\mathsf{ap_{pre}}$ **of** $\mathsf{pre}$ **computes preimages).** *Let* $\mathsf{f} : \mathsf{X} \rightsquigarrow_{\perp} \mathsf{Y}$*. For all* $\mathsf{A} \subseteq \mathsf{X}$ *and* $\mathsf{B} \subseteq \mathsf{Y}$, $\mathsf{ap_{pre}}\ (\mathsf{pre\ f\ A})\ \mathsf{B} \equiv \mathsf{preimage_{\perp}\ f\ A\ B}$.

### 4.3 The Preimage Arrow

If we define the **preimage arrow** type constructor as

$$\mathsf{X} \rightsquigarrow_{\mathsf{pre}} \mathsf{Y} ::= \mathsf{Set\ X} \to (\mathsf{X} \rightharpoonup_{\mathsf{pre}} \mathsf{Y}) \tag{23}$$

then we already have a lift $\mathsf{lift_{pre}} : (\mathsf{X} \rightsquigarrow_{\perp} \mathsf{Y}) \to (\mathsf{X} \rightsquigarrow_{\mathsf{pre}} \mathsf{Y})$ from the bottom arrow to the preimage arrow: $\mathsf{pre}$. If $\mathsf{lift_{pre}}$ is $\mathsf{pre}$, then by Theorem 3, lifted bottom arrow computations compute correct preimages, exactly as we should expect them to.

Fig. 5 defines the preimage arrow in terms of the preimage function operations in Fig. 4. For these definitions to make $\mathsf{lift_{pre}}$ a homomorphism, preimage arrow equivalence must mean "computes the same preimages."

**Definition 5 (preimage arrow equivalence).** *Two preimage arrow computations* $\mathsf{h_1} : \mathsf{X} \rightsquigarrow_{\mathsf{pre}} \mathsf{Y}$ *and* $\mathsf{h_2} : \mathsf{X} \rightsquigarrow_{\mathsf{pre}} \mathsf{Y}$ *are equivalent, or* $\mathsf{h_1} \equiv \mathsf{h_2}$*, when* $\mathsf{ap_{pre}}\ (\mathsf{h_1\ A})\ \mathsf{B} \equiv \mathsf{ap_{pre}}\ (\mathsf{h_2\ A})\ \mathsf{B}$ *for all* $\mathsf{A} \subseteq \mathsf{X}$ *and* $\mathsf{B} \subseteq \mathsf{Y}$.

**Theorem 4 (preimage arrow correctness).** $\mathsf{lift_{pre}}$ *is a homomorphism.*

**Corollary 1 (semantic correctness).** *For all* $e$, $[\![e]\!]_{\mathsf{pre}} \equiv \mathsf{lift_{pre}}\ [\![e]\!]_{\perp}$.

In other words, $[\![e]\!]_{\mathsf{pre}}$ always computes correct preimages under $[\![e]\!]_{\perp}$.

Inhabitants of type $\mathsf{X} \rightsquigarrow_{\mathsf{pre}} \mathsf{Y}$ do not always behave intuitively; e.g.

$$
\begin{aligned}
&\mathsf{unruly} : \mathsf{Bool} \rightsquigarrow_{\mathsf{pre}} \mathsf{Bool} \\
&\mathsf{unruly\ A} := \langle \mathsf{Bool} \backslash \mathsf{A}, \lambda \mathsf{B}.\, \mathsf{B} \rangle
\end{aligned}
\tag{24}
$$

So $\mathsf{ap_{pre}}\ (\mathsf{unruly}\ \{\mathsf{true}\})\ \{\mathsf{false}\} = \{\mathsf{false}\} \cap (\mathsf{Bool} \backslash \{\mathsf{true}\}) = \{\mathsf{false}\}$—a "preimage" that does not even intersect the given domain $\{\mathsf{true}\}$. Other examples show that preimage computations are not necessarily monotone, and lack other desirable properties. Those with desirable properties obey the following law.

**Definition 6 (preimage arrow law).** *Let* $\mathsf{h} : \mathsf{X} \rightsquigarrow_{\mathsf{pre}} \mathsf{Y}$*. If there exists an* $\mathsf{f} : \mathsf{X} \rightsquigarrow_{\perp} \mathsf{Y}$ *such that* $\mathsf{h} \equiv \mathsf{lift_{pre}}\ \mathsf{f}$*, then* $\mathsf{h}$ *obeys the* ***preimage arrow law****.*

By homomorphism of $\mathsf{lift_{pre}}$, preimage arrow combinators preserve the preimage arrow law. From here on, we assume all $\mathsf{h} : \mathsf{X} \rightsquigarrow_{\mathsf{pre}} \mathsf{Y}$ obey it. By Definition 6, $\mathsf{lift_{pre}}$ has a right inverse; by Theorem 1, the arrow laws hold.

## 5 The Bottom* and Preimage* Arrows

This section lifts the prior semantics to recursive, probabilistic programs.

We have defined the top of our roadmap:

$$\begin{array}{ccc} \mathsf{X} \rightsquigarrow_{\bot} \mathsf{Y} & \xrightarrow{\mathsf{lift}_{\mathsf{pre}}} & \mathsf{X} \rightsquigarrow_{\mathsf{pre}} \mathsf{Y} \\ \eta_{\bot^*} \downarrow & & \downarrow \eta_{\mathsf{pre}^*} \\ \mathsf{X} \rightsquigarrow_{\bot^*} \mathsf{Y} & \xrightarrow[\mathsf{lift}_{\mathsf{pre}^*}]{} & \mathsf{X} \rightsquigarrow_{\mathsf{pre}^*} \mathsf{Y} \end{array} \tag{25}$$

so that $\mathsf{lift}_{\mathsf{pre}}$ is a homomorphism. Now we move down each side and connect the bottom, in a way that makes every morphism a homomorphism.

Probabilistic functions that may not terminate, but terminate with probability 1, are common. For example, suppose $\mathsf{random}$ retrieves numbers in $[0,1]$ from an implicit random source. The following probabilistic function defines the well-known geometric distribution by counting the number of times $\mathsf{random} < \mathsf{p}$:

$$\mathsf{geometric\ p} \ := \ \mathsf{if\ random} < \mathsf{p\ then}\ 0\ \mathsf{else}\ 1 + \mathsf{geometric\ p} \tag{26}$$

For any $\mathsf{p} > 0$, $\mathsf{geometric\ p}$ may not terminate, but the probability of not terminating (i.e. always taking the "else" branch) is $(1-\mathsf{p}) \cdot (1-\mathsf{p}) \cdot (1-\mathsf{p}) \cdot \dots = 0$.

Suppose we interpret $\mathsf{geometric\ p}$ as $\mathsf{h} : \mathsf{R} \rightsquigarrow_{\mathsf{pre}} \mathbb{N}$, a preimage arrow computation from random sources to $\mathbb{N}$, and we have a probability measure $\mathsf{P} : \mathsf{Set\ R} \to [0,1]$. The probability of $\mathsf{N} \subseteq \mathbb{N}$ is $\mathsf{P}\ (\mathsf{ap}_{\mathsf{pre}}\ (\mathsf{h\ R})\ \mathsf{N})$. To compute this, we must

- Ensure each $\mathsf{r} \in \mathsf{R}$ contains enough random numbers.
- Determine how $\mathsf{random}$ indexes numbers in $\mathsf{r}$.
- Ensure $\mathsf{ap}_{\mathsf{pre}}\ (\mathsf{h\ R})\ \mathsf{N}$ terminates even though there are random sources in $\mathsf{R}$ for which $\mathsf{geometric\ p}$ does not terminate.

The last task is the most difficult, but doing the first two will provide structure that makes it much easier.

### 5.1 Threading and Indexing

We need bottom and preimage arrows that thread a random source. To ensure random sources contain enough numbers, they should be infinite.

In a pure $\lambda$-calculus, random sources are typically infinite streams, threaded monadically: each computation receives and produces a random source. A little-used alternative is for the random source to be an infinite tree, threaded applicatively: each computation receives, but does not produce, a random source. Combinators split the tree and pass subtrees to subcomputations.

With either alternative, for arrows, the resulting definitions are large, conceptually difficult, and hard to manipulate. Fortunately, it is relatively easy to assign each subcomputation a unique index into a tree-shaped random source and pass the random source unchanged. For this, we need an indexing scheme.

**Definition 7 (binary indexing scheme).** *Let* $\mathsf{J}$ *be the set of finite lists of* $\mathsf{Bool}$. *Define* $\mathsf{j}_0 := \langle\rangle$ *as the root node's index, and* $\mathsf{left} : \mathsf{J} \to \mathsf{J}$*;* $\mathsf{left\ j} := \langle \mathsf{true}, \mathsf{j} \rangle$ *and* $\mathsf{right} : \mathsf{J} \to \mathsf{J}$*;* $\mathsf{right\ j} := \langle \mathsf{false}, \mathsf{j} \rangle$ *to construct left and right child indexes.*

$$
\begin{aligned}
\mathsf{AStore}\ \mathsf{s}\ (\mathsf{x} \rightsquigarrow_{\mathsf{a}} \mathsf{y}) \ &::=\ \mathsf{J} \to (\langle \mathsf{s}, \mathsf{x} \rangle \rightsquigarrow_{\mathsf{a}} \mathsf{y}) \\
\mathsf{x} \rightsquigarrow_{\mathsf{a}^*} \mathsf{y} \ &::=\ \mathsf{AStore}\ \mathsf{s}\ (\mathsf{x} \rightsquigarrow_{\mathsf{a}} \mathsf{y}) \\
\\
\mathsf{arr}_{\mathsf{a}^*} \ &:=\ \eta_{\mathsf{a}^*} \circ \mathsf{arr}_{\mathsf{a}} \\
(\mathsf{k}_1 \ggg_{\mathsf{a}^*} \mathsf{k}_2)\ \mathsf{j} \ &:=\ (\mathsf{arr}_{\mathsf{a}}\ \mathsf{fst}\ \&\&\&_{\mathsf{a}}\ \mathsf{k}_1\ (\mathsf{left}\ \mathsf{j})) \ggg_{\mathsf{a}} \mathsf{k}_2\ (\mathsf{right}\ \mathsf{j}) \\
(\mathsf{k}_1\ \&\&\&_{\mathsf{a}^*}\ \mathsf{k}_2)\ \mathsf{j} \ &:=\ \mathsf{k}_1\ (\mathsf{left}\ \mathsf{j})\ \&\&\&_{\mathsf{a}}\ \mathsf{k}_2\ (\mathsf{right}\ \mathsf{j})
\end{aligned}
$$

$$
\begin{aligned}
&\mathsf{ifte}_{\mathsf{a}^*}\ \mathsf{k}_1\ \mathsf{k}_2\ \mathsf{k}_3\ \mathsf{j} \ := \\
&\quad \mathsf{ifte}_{\mathsf{a}}\ (\mathsf{k}_1\ (\mathsf{left}\ \mathsf{j})) \\
&\qquad\quad (\mathsf{k}_2\ (\mathsf{left}\ (\mathsf{right}\ \mathsf{j}))) \\
&\qquad\quad (\mathsf{k}_3\ (\mathsf{right}\ (\mathsf{right}\ \mathsf{j}))) \\
&\mathsf{lazy}_{\mathsf{a}^*}\ \mathsf{k}\ \mathsf{j} \ :=\ \mathsf{lazy}_{\mathsf{a}}\ \lambda 0.\, \mathsf{k}\ 0\ \mathsf{j} \\
&\eta_{\mathsf{a}^*}\ \mathsf{f}\ \mathsf{j} \ :=\ \mathsf{arr}_{\mathsf{a}}\ \mathsf{snd} \ggg_{\mathsf{a}} \mathsf{f}
\end{aligned}
$$

Fig. 6: AStore (associative store) arrow transformer definitions.

We define random-source-threading variants of both the bottom and preimage arrows at the same time by defining an **arrow transformer**: an arrow parameterized on another arrow. The AStore arrow transformer type constructor takes a store type $\mathsf{s}$ and an arrow $\mathsf{x} \rightsquigarrow_{\mathsf{a}} \mathsf{y}$:

$$\mathsf{AStore}\ \mathsf{s}\ (\mathsf{x} \rightsquigarrow_{\mathsf{a}} \mathsf{y}) \ ::=\ \mathsf{J} \to (\langle \mathsf{s}, \mathsf{x} \rangle \rightsquigarrow_{\mathsf{a}} \mathsf{y}) \tag{27}$$

Reading the type, we see that computations receive an index $\mathsf{j} \in \mathsf{J}$ and produce a computation that receives a store as well as an $\mathsf{x}$. Lifting extracts the $\mathsf{x}$ from the input pair and sends it on to the original computation, ignoring $\mathsf{j}$:

$$
\begin{aligned}
&\eta_{\mathsf{a}^*} : (\mathsf{x} \rightsquigarrow_{\mathsf{a}} \mathsf{y}) \to \mathsf{AStore}\ \mathsf{s}\ (\mathsf{x} \rightsquigarrow_{\mathsf{a}} \mathsf{y}) \\
&\eta_{\mathsf{a}^*}\ \mathsf{f}\ \mathsf{j} \ :=\ \mathsf{arr}_{\mathsf{a}}\ \mathsf{snd} \ggg_{\mathsf{a}} \mathsf{f}
\end{aligned} \tag{28}
$$

Fig. 6 defines the remaining combinators. Each subcomputation receives left j, right j, or some other unique binary index. We thus think of programs interpreted as AStore arrows as being completely unrolled into an infinite binary tree, with each expression labeled with its tree index.

### 5.2 Recursive, Probabilistic Programs

To interpret probabilistic programs, we put infinite random trees in the store.

Of all the ways to represent infinite binary trees whose nodes are labeled with values in $[0,1]$, the way most compatible with measure theory is to flatten them into vectors of $[0,1]$ indexed by $\mathsf{J}$. The set of all such vectors is $[0,1]^{\mathsf{J}}$.

**Definition 8 (random source).** *Define* $\mathsf{R} := [0,1]^{\mathsf{J}}$*, the set of infinite binary trees whose node labels are in* $[0,1]$*. A **random source** is any* $\mathsf{r} \in \mathsf{R}$*.*

To interpret recursive programs, we need to ensure termination. One ultimately implementable way is to have the store dictate which branch of each conditional, if any, is taken. If the store dictates that all but finitely many branches cannot be taken, well-defined programs must terminate (see Definition 3).

**Definition 9 (branch trace).** *A **branch trace** is any* $\mathsf{t} \in (\mathsf{Bool}_\perp)^{\mathsf{J}}$ *such that* $\mathsf{t}\ \mathsf{j} = \mathsf{true}$ *or* $\mathsf{t}\ \mathsf{j} = \mathsf{false}$ *for no more than finitely many* $\mathsf{j} \in \mathsf{J}$*.*

*Let* $\mathsf{T} \subset (\mathsf{Bool}_\perp)^{\mathsf{J}}$ *be the set of all branch traces.*

Let $\mathsf{X} \leadsto_{\mathsf{a}^*} \mathsf{Y} ::= \mathsf{AStore}\ \langle \mathsf{R}, \mathsf{T} \rangle\ (\mathsf{X} \leadsto_{\mathsf{a}} \mathsf{Y})$ denote the AStore arrow type that threads both random sources and branch traces through another arrow a. Thus, the type constructors for the **bottom*** and **preimage*** arrows are

$$\begin{aligned} \mathsf{X} \leadsto_{\perp^*} \mathsf{Y} &::= \mathsf{AStore}\ \langle \mathsf{R}, \mathsf{T} \rangle\ (\mathsf{X} \leadsto_{\perp} \mathsf{Y}) \\ \mathsf{X} \leadsto_{\mathsf{pre}^*} \mathsf{Y} &::= \mathsf{AStore}\ \langle \mathsf{R}, \mathsf{T} \rangle\ (\mathsf{X} \leadsto_{\mathsf{pre}} \mathsf{Y}) \end{aligned} \tag{29}$$

For probabilistic programs, we define a combinator $\mathsf{random}_{\mathsf{a}^*}$ that returns the number at its tree index in the random source, and extend $[\![\cdot]\!]_{\mathsf{a}^*}$ for arrows $\mathsf{a}^*$:

$$\begin{aligned} &\mathsf{random}_{\mathsf{a}^*} : \mathsf{X} \leadsto_{\mathsf{a}^*} [0,1] & [\![\mathsf{random}]\!]_{\mathsf{a}^*} :\equiv \mathsf{random}_{\mathsf{a}^*} \\ &\mathsf{random}_{\mathsf{a}^*}\ \mathsf{j} := \mathsf{arr_a\ fst} \ggg_{\mathsf{a}} \mathsf{arr_a\ fst} \ggg_{\mathsf{a}} \mathsf{arr_a}\ (\pi\ \mathsf{j}) & \end{aligned} \tag{30}$$

where $\pi : \mathsf{J} \to \mathsf{X}^{\mathsf{J}} \to \mathsf{X}$, defined by $\pi\ \mathsf{j}\ \mathsf{f} := \mathsf{f}\ \mathsf{j}$, produces projection functions.

For recursive programs, we define a combinator that reads branch traces, and a new if-then-else combinator that yields $\perp$ when its test expression does not agree with the branch trace at its tree index:

$$\begin{aligned} &\mathsf{branch}_{\mathsf{a}^*} : \mathsf{X} \leadsto_{\mathsf{a}^*} \mathsf{Bool} \\ &\mathsf{branch}_{\mathsf{a}^*}\ \mathsf{j} := \mathsf{arr_a\ fst} \ggg_{\mathsf{a}} \mathsf{arr_a\ snd} \ggg_{\mathsf{a}} \mathsf{arr_a}\ (\pi\ \mathsf{j}) \\ \\ &\mathsf{ifte}^{\Downarrow}_{\mathsf{a}^*} : (\mathsf{x} \leadsto_{\mathsf{a}^*} \mathsf{Bool}) \to (\mathsf{x} \leadsto_{\mathsf{a}^*} \mathsf{y}) \to (\mathsf{x} \leadsto_{\mathsf{a}^*} \mathsf{y}) \to (\mathsf{x} \leadsto_{\mathsf{a}^*} \mathsf{y}) \\ &\mathsf{ifte}^{\Downarrow}_{\mathsf{a}^*}\ \mathsf{k}_1\ \mathsf{k}_2\ \mathsf{k}_3\ \mathsf{j} := \mathsf{ifte_a}\ ((\mathsf{k}_1\ (\mathsf{left\ j}) \ \&\&\&_{\mathsf{a}}\ \mathsf{branch}_{\mathsf{a}^*}\ \mathsf{j}) \ggg_{\mathsf{a}} \mathsf{arr_a\ agrees}) \\ &\qquad\qquad\qquad\qquad (\mathsf{k}_2\ (\mathsf{left\ (right\ j)})) \\ &\qquad\qquad\qquad\qquad (\mathsf{k}_3\ (\mathsf{right\ (right\ j)})) \end{aligned} \tag{31}$$

where $\mathsf{agrees}\ \langle \mathsf{b}_1, \mathsf{b}_2 \rangle := \mathsf{if}\ \mathsf{b}_1 = \mathsf{b}_2\ \mathsf{then}\ \mathsf{b}_1\ \mathsf{else}\ \perp$. We define a new semantic function $[\![\cdot]\!]^{\Downarrow}_{\mathsf{a}^*}$ by replacing the if rule in $[\![\cdot]\!]_{\mathsf{a}^*}$:

$$[\![\mathsf{if}\ e_c\ \mathsf{then}\ e_t\ \mathsf{else}\ e_f]\!]^{\Downarrow}_{\mathsf{a}^*} :\equiv \mathsf{ifte}^{\Downarrow}_{\mathsf{a}^*}\ [\![e_c]\!]^{\Downarrow}_{\mathsf{a}^*}\ (\mathsf{lazy}_{\mathsf{a}^*}\ \lambda 0.\, [\![\mathsf{e_t}]\!]^{\Downarrow}_{\mathsf{a}^*})\ (\mathsf{lazy}_{\mathsf{a}^*}\ \lambda 0.\, [\![\mathsf{e_f}]\!]^{\Downarrow}_{\mathsf{a}^*}) \tag{32}$$

Suppose $\mathsf{f} := ([\![p]\!]^{\Downarrow}_{\perp^*}\ \mathsf{j}_0) : \mathsf{X}' \leadsto_{\perp} \mathsf{Y}$ and $\mathsf{h} := ([\![p]\!]^{\Downarrow}_{\mathsf{pre}^*}\ \mathsf{j}_0) : \mathsf{X}' \leadsto_{\mathsf{pre}} \mathsf{Y}$, where $\mathsf{X}' = (\mathsf{R} \times \mathsf{T}) \times \mathsf{X}$. For each $\langle \langle \mathsf{r}, \mathsf{t} \rangle, \mathsf{a} \rangle \in \mathsf{X}'$, we assume that only r is chosen randomly. Thus, the probability of $\mathsf{B} \subseteq \mathsf{Y}$ is

$$\begin{aligned} &\mathsf{P}\ (\mathsf{image}\ (\mathsf{fst} \circ \mathsf{fst})\ (\mathsf{preimage}_{\perp}\ \mathsf{f}\ \mathsf{X}'\ \mathsf{B})) \\ &\quad = \mathsf{P}\ (\mathsf{image}\ (\mathsf{fst} \circ \mathsf{fst})\ (\mathsf{ap_{pre}}\ (\mathsf{h}\ \mathsf{X}')\ \mathsf{B})) \end{aligned} \tag{33}$$

if f and h always terminate and $[\![\cdot]\!]^{\Downarrow}_{\mathsf{pre}^*}$ is correct with respect to $[\![\cdot]\!]^{\Downarrow}_{\perp^*}$.

### 5.3 Correctness and Termination

The proofs in this section require AStore arrow equivalence to be a little coarser.

**Definition 10** (AStore **arrow equivalence**). *Two* AStore *arrow computations* $\mathsf{k}_1$ *and* $\mathsf{k}_2$ *are equivalent, or* $\mathsf{k}_1 \equiv \mathsf{k}_2$*, when* $\mathsf{k}_1\ \mathsf{j} \equiv \mathsf{k}_2\ \mathsf{j}$ *for all* $\mathsf{j} \in \mathsf{J}$.

Proving $[\![\cdot]\!]_{\bot^*}$ and $[\![\cdot]\!]_{\mathsf{pre}^*}$ correct with respect to $[\![\cdot]\!]_{\bot}$ and $[\![\cdot]\!]_{\mathsf{pre}}$, for programs without $\mathsf{random}$, only requires proving $\eta_{\mathsf{a}^*}$ homomorphic, using the arrow laws.

**Theorem 5 (pure AStore arrow correctness).** *$\eta_{\mathsf{a}^*}$ is a homomorphism.*

**Corollary 2 (pure semantic correctness).** *For all pure e, $[\![e]\!]_{\mathsf{a}^*} \equiv \eta_{\mathsf{a}^*}\ [\![e]\!]_{\mathsf{a}}$.*

We use a homomorphic lift to prove $[\![\cdot]\!]^{\Downarrow}_{\mathsf{pre}^*}$ correct with respect to $[\![\cdot]\!]^{\Downarrow}_{\bot^*}$ If we define it in terms of $\mathsf{lift_b} : (\mathsf{x} \leadsto_{\mathsf{a}} \mathsf{y}) \to (\mathsf{x} \leadsto_{\mathsf{b}} \mathsf{y})$ as

$$\begin{aligned} &\mathsf{lift}_{\mathsf{b}^*} : (\mathsf{x} \leadsto_{\mathsf{a}^*} \mathsf{y}) \to (\mathsf{x} \leadsto_{\mathsf{b}^*} \mathsf{y}) \\ &\mathsf{lift}_{\mathsf{b}^*}\ \mathsf{f}\ \mathsf{j}\ := \ \mathsf{lift_b}\ (\mathsf{f}\ \mathsf{j}) \end{aligned} \tag{34}$$

then we need only use the fact that $\mathsf{a}$ and $\mathsf{b}$ are arrows to prove the following.

**Theorem 6 (effectful AStore arrow correctness).** *If $\mathsf{lift_b}$ is an arrow homomorphism from $\mathsf{a}$ to $\mathsf{b}$, then $\mathsf{lift}_{\mathsf{b}^*}$ is an arrow homomorphism from $\mathsf{a}^*$ to $\mathsf{b}^*$.*

**Corollary 3 (effectful semantic correctness).** *For all e, $[\![e]\!]_{\mathsf{pre}^*} \equiv \mathsf{lift}_{\mathsf{pre}^*}\ [\![e]\!]_{\bot^*}$ and $[\![e]\!]^{\Downarrow}_{\mathsf{pre}^*} \equiv \mathsf{lift}_{\mathsf{pre}^*}\ [\![e]\!]^{\Downarrow}_{\bot^*}$.*

For termination, we need to define the largest domain on which $[\![e]\!]^{\Downarrow}_{\mathsf{a}^*}$ and $[\![e]\!]_{\mathsf{a}^*}$ computations should agree.

**Definition 11 (maximal domain).** *Let $\mathsf{f} : \mathsf{X} \leadsto_{\bot^*} \mathsf{Y}$. Its **maximal domain** is the largest $\mathsf{A}^* \subseteq (\mathsf{R} \times \mathsf{T}) \times \mathsf{X}$ for which $\mathsf{A}^* = \{\mathsf{a} \in \mathsf{A}^* \mid \mathsf{f}\ \mathsf{j}_0\ \mathsf{a} \neq \bot\}$.*

Because $\mathsf{f}\ \mathsf{j}_0\ \mathsf{a} \neq \bot$ implies termination, all inputs in $\mathsf{A}^*$ are terminating.

**Theorem 7 (correct termination everywhere).** *Let $[\![e]\!]^{\Downarrow}_{\bot^*} : \mathsf{X} \leadsto_{\bot^*} \mathsf{Y}$ have maximal domain $\mathsf{A}^*$, and $\mathsf{X}' := (\mathsf{R} \times \mathsf{T}) \times \mathsf{X}$. For all $\mathsf{a} \in \mathsf{X}'$, $\mathsf{A} \subseteq \mathsf{X}'$ and $\mathsf{B} \subseteq \mathsf{Y}$,*

$$\begin{aligned} [\![e]\!]^{\Downarrow}_{\bot^*}\ \mathsf{j}_0\ \mathsf{a} \quad &= \ \mathsf{if}\ \mathsf{a} \in \mathsf{A}^*\ \mathsf{then}\ [\![e]\!]_{\bot^*}\ \mathsf{j}_0\ \mathsf{a}\ \mathsf{else}\ \bot \\ \mathsf{ap_{pre}}\ ([\![e]\!]^{\Downarrow}_{\mathsf{pre}^*}\ \mathsf{j}_0\ \mathsf{A})\ \mathsf{B} \ &= \ \mathsf{ap_{pre}}\ ([\![e]\!]_{\mathsf{pre}^*}\ \mathsf{j}_0\ (\mathsf{A} \cap \mathsf{A}^*))\ \mathsf{B} \end{aligned} \tag{35}$$

In other words, $[\![\cdot]\!]^{\Downarrow}_{\mathsf{pre}^*}$ computations always terminate, and the sets they yield are correct preimages.

## 6 Abstract Semantics

This section derives a sound, implementable abstract semantics. Most preimages of uncountable sets are uncomputable. We therefore define a semantics for approximate preimage computation by

1. Choosing abstract set types that can be finitely represented, and operations that overapproximate concrete set operations.
2. Replacing concrete set types and operations with abstract set types and operations in the definitions of the preimage and preimage* arrows.
3. Proving termination, soundness, and other desirable properties.

In a sense, this is typical abstract interpretation. However, not having a fixpoint operator in the language means there is no abstract fixpoint to compute, and abstract preimage arrow computations actually apply functions.

### 6.1 Abstract Sets

We use the abstract domain of rectangles with an atypical extension to represent rectangles of $\mathsf{X}^\mathsf{J}$ (i.e. infinite binary trees of $\mathsf{X}$).

**Definition 12 (rectangular sets).** *For a type* $\mathsf{X}$ *of language values,* $\mathsf{Rect\ X}$ *denotes the type of* ***rectangular sets*** *of* $\mathsf{X}$*: a bounded lattice of sets in* $\mathsf{Set\ X}$ *ordered by* $(\subseteq)$*; i.e. it contains* $\varnothing$ *and* $\mathsf{X}$*, and is closed under meet* $(\cap)$ *and join* $(\sqcup)$*. Rectangles of cartesian products are defined by*

$$\mathsf{Rect}\ \langle \mathsf{X}_1, \mathsf{X}_2 \rangle \ ::= \ \{\mathsf{A}_1 \times \mathsf{A}_2 \mid \mathsf{A}_1 : \mathsf{Rect\ X}_1, \mathsf{A}_2 : \mathsf{Rect\ X}_2\} \tag{36}$$

*Rectangles of infinite binary trees (i.e. products indexed by* $\mathsf{J}$*) are defined by*

$$\mathsf{Rect\ X}^\mathsf{J} \ ::= \bigcup_{\mathsf{J}' \subset \mathsf{J}\ \mathit{finite}} \left\{ \textstyle\prod_{\mathsf{j} \in \mathsf{J}} \mathsf{A}_\mathsf{j} \ \middle|\ \mathsf{A}_\mathsf{j} : \mathsf{Rect\ X},\ \mathsf{j} \notin \mathsf{J}' \iff \mathsf{A}_\mathsf{j} = \mathsf{X} \right\} \tag{37}$$

*i.e. for* $\mathsf{A} : \mathsf{Rect\ X}^\mathsf{J}$*, only finitely many axes of* $\mathsf{A}$ *are proper subsets of* $\mathsf{X}$*. Joins of products are defined by*

$$(\mathsf{A}_1 \times \mathsf{A}_2) \sqcup (\mathsf{B}_1 \times \mathsf{B}_2) \ = \ (\mathsf{A}_1 \sqcup \mathsf{B}_1) \times (\mathsf{A}_2 \sqcup \mathsf{B}_2) \tag{38}$$

$$(\textstyle\prod_{\mathsf{j} \in \mathsf{J}} \mathsf{A}_\mathsf{j}) \sqcup (\prod_{\mathsf{j} \in \mathsf{J}} \mathsf{B}_\mathsf{j}) \ = \ \prod_{\mathsf{j} \in \mathsf{J}} (\mathsf{A}_\mathsf{j} \sqcup \mathsf{B}_\mathsf{j}) \tag{39}$$

The lattice properties imply that $(\sqcup)$ overapproximates $(\cup)$; i.e. $\mathsf{A} \cup \mathsf{B} \subseteq \mathsf{A} \sqcup \mathsf{B}$. For non-product types $\mathsf{X}$, $\mathsf{Rect\ X}$ may be any bounded sublattice of $\mathsf{Set\ X}$. Interpreting conditionals requires $\{\mathsf{true}\}$ and $\{\mathsf{false}\}$; thus $\mathsf{Rect\ Bool} ::= \mathsf{Set\ Bool}$.

Intervals in ordered spaces can be implemented as pairs of endpoints. Products in $\mathsf{Rect}\ \langle \mathsf{X}_1, \mathsf{X}_2 \rangle$ can be implemented as pairs of type $\langle \mathsf{Rect\ X}_1, \mathsf{Rect\ X}_2 \rangle$. By (37), products in $\mathsf{Rect\ X}^\mathsf{J}$ have only finitely many axes that are proper subsets of $\mathsf{X}$, so they can be implemented as *finite* binary trees. All operations on products proceed by simple structural recursion.

### 6.2 Abstract Arrows

To define the abstract preimage arrow, we start by defining abstract preimage functions, by replacing set types in $(\rightharpoonup_{\mathsf{pre}})$ with abstract set types:

$$\mathsf{X} \rightharpoonup_{\widehat{\mathsf{pre}}} \mathsf{Y} ::= \langle \mathsf{Rect\ Y}, \mathsf{Rect\ Y} \to \mathsf{Rect\ X} \rangle \tag{40}$$

Fig. 7a defines the necessary operations on abstract preimage functions by replacing set operations with *abstract* set operations—except for $\langle \cdot, \cdot \rangle_{\widehat{\mathsf{pre}}}$, which is greatly simplified by the fact that $\mathsf{preimage}$ distributes over pairing and products (12). (Compare Fig. 4.) Similarly, Fig. 7b defines the abstract preimage arrow by replacing preimage function types and operations in the preimage arrow's definition with *abstract* preimage function types and operations. (Compare Fig. 5.) The lift $\mathsf{arr}_{\widehat{\mathsf{pre}}} : (\mathsf{X} \to \mathsf{Y}) \to (\mathsf{X} \rightsquigarrow_{\widehat{\mathsf{pre}}} \mathsf{Y})$ exists, but $\mathsf{arr}_{\widehat{\mathsf{pre}}}\ \mathsf{f}$ is not always unique (because by definition, $\mathsf{Rect\ X}^\mathsf{J}$ is an incomplete lattice) nor computable.

$$X \rightharpoonup_{\widehat{pre}} Y ::= \langle \mathrm{Rect}\ Y, \mathrm{Rect}\ Y \to \mathrm{Rect}\ X \rangle$$

$$\varnothing_{\widehat{pre}} := \langle \varnothing, \lambda B.\, \varnothing \rangle$$

$$\begin{aligned}
&\mathrm{ap}_{\widehat{pre}} : (X \rightharpoonup_{\widehat{pre}} Y) \to \mathrm{Rect}\ Y \to \mathrm{Rect}\ X \\
&\mathrm{ap}_{\widehat{pre}}\ \langle Y', p \rangle\ B := p\ (B \cap Y')
\end{aligned}$$

$$\begin{aligned}
&\mathrm{range}_{\widehat{pre}} : (X \rightharpoonup_{\widehat{pre}} Y) \to \mathrm{Rect}\ Y \\
&\mathrm{range}_{\widehat{pre}}\ \langle Y', p \rangle := Y'
\end{aligned}$$

$$\begin{aligned}
&\langle \cdot, \cdot \rangle_{\widehat{pre}} : (X \rightharpoonup_{\widehat{pre}} Y_1) \to (X \rightharpoonup_{\widehat{pre}} Y_2) \\
&\qquad \to (X \rightharpoonup_{\widehat{pre}} \langle Y_1, Y_2 \rangle) \\
&\langle \langle Y'_1, p_1 \rangle, \langle Y'_2, p_2 \rangle \rangle_{\widehat{pre}} := \\
&\quad \langle Y'_1 \times Y'_2, \lambda B.\, p_1\ (\mathrm{proj}_1\ B) \cap p_2\ (\mathrm{proj}_2\ B) \rangle
\end{aligned}$$

$$\begin{aligned}
&(\circ_{\widehat{pre}}) : (Y \rightharpoonup_{\widehat{pre}} Z) \to (X \rightharpoonup_{\widehat{pre}} Y) \to (X \rightharpoonup_{\widehat{pre}} Z) \\
&\langle Z', p_2 \rangle \circ_{\widehat{pre}} h_1 := \langle Z', \lambda C.\, \mathrm{ap}_{\widehat{pre}}\ h_1\ (p_2\ C) \rangle
\end{aligned}$$

$$\begin{aligned}
&(\cup_{\widehat{pre}}) : (X \rightharpoonup_{\widehat{pre}} Y) \to (X \rightharpoonup_{\widehat{pre}} Y) \to (X \rightharpoonup_{\widehat{pre}} Y) \\
&\langle Y'_1, p_1 \rangle \cup_{\widehat{pre}} \langle Y'_2, p_2 \rangle := \\
&\quad \langle Y'_1 \sqcup Y'_2, \lambda B.\, \mathrm{ap}_{\widehat{pre}}\ \langle Y'_1, p_1 \rangle\ B \sqcup \mathrm{ap}_{\widehat{pre}}\ \langle Y'_2, p_2 \rangle\ B \rangle
\end{aligned}$$

(a) Definitions for abstract preimage functions, which compute rectangular covers.

$$X \rightsquigarrow_{\widehat{pre}} Y ::= \mathrm{Rect}\ X \to (X \rightharpoonup_{\widehat{pre}} Y)$$

$$\begin{aligned}
(h_1 \ggg_{\widehat{pre}} h_2)\ A := \ &\mathrm{let}\ h'_1 := h_1\ A \\
&\qquad h'_2 := h_2\ (\mathrm{range}_{\widehat{pre}}\ h'_1) \\
&\mathrm{in}\ h'_2 \circ_{\widehat{pre}} h'_1
\end{aligned}$$

$$(h_1 \ \&\&\&_{\widehat{pre}}\ h_2)\ A := \langle h_1\ A, h_2\ A \rangle_{\widehat{pre}}$$

$$\begin{aligned}
&\mathrm{ifte}_{\widehat{pre}}\ h_1\ h_2\ h_3\ A := \\
&\quad \mathrm{let}\ h'_1 := h_1\ A \\
&\qquad\quad h'_2 := h_2\ (\mathrm{ap}_{\widehat{pre}}\ h'_1\ \{\mathrm{true}\}) \\
&\qquad\quad h'_3 := h_3\ (\mathrm{ap}_{\widehat{pre}}\ h'_1\ \{\mathrm{false}\}) \\
&\quad \mathrm{in}\ h'_2 \cup_{\widehat{pre}} h'_3
\end{aligned}$$

$$\mathrm{lazy}_{\widehat{pre}}\ h\ A := \mathrm{if}\ A = \varnothing\ \mathrm{then}\ \varnothing_{\widehat{pre}}\ \mathrm{else}\ h\ 0\ A$$

(b) Abstract preimage arrow, defined using abstract preimage functions.

$$\begin{aligned}
\mathrm{id}_{\widehat{pre}}\ A &:= \langle A, \lambda B.\, B \rangle \\
\mathrm{fst}_{\widehat{pre}}\ A &:= \langle \mathrm{proj}_1\ A, \mathrm{unproj}_1\ A \rangle \\
\mathrm{snd}_{\widehat{pre}}\ A &:= \langle \mathrm{proj}_2\ A, \mathrm{unproj}_2\ A \rangle
\end{aligned}$$

$$\begin{aligned}
&\mathrm{proj}_1 := \mathrm{image}\ \mathrm{fst} \\
&\mathrm{proj}_2 := \mathrm{image}\ \mathrm{snd} \\
&\mathrm{unproj}_1\ A\ B := A \cap (B \times \mathrm{proj}_2\ A) \\
&\mathrm{unproj}_2\ A\ B := A \cap (\mathrm{proj}_1\ A \times B)
\end{aligned}$$

$$\begin{aligned}
\mathrm{const}_{\widehat{pre}}\ b\ A &:= \langle \{b\}, \lambda B.\, \mathrm{if}\ B = \varnothing\ \mathrm{then}\ \varnothing\ \mathrm{else}\ A \rangle \\
\pi_{\widehat{pre}}\ j\ A &:= \langle \mathrm{proj}\ j\ A, \mathrm{unproj}\ j\ A \rangle
\end{aligned}$$

$$\begin{aligned}
&\mathrm{proj} : J \to \mathrm{Set}\ X^J \to \mathrm{Set}\ X \\
&\mathrm{proj}\ j\ A := \mathrm{image}\ (\pi\ j)\ A \\
&\mathrm{unproj} : J \to \mathrm{Set}\ X^J \to \mathrm{Set}\ X \to \mathrm{Set}\ X^J \\
&\mathrm{unproj}\ j\ A\ B := A \cap \textstyle\prod_{i \in J}\ \mathrm{if}\ j = i\ \mathrm{then}\ B\ \mathrm{else}\ \mathrm{proj}\ j\ A
\end{aligned}$$

(c) Explicit instances of $\mathrm{arr}_{\widehat{pre}}\ f$ (e.g. $\mathrm{arr}_{\widehat{pre}}\ \mathrm{id}$) needed to interpret probabilistic programs.

$$X \rightsquigarrow_{\widehat{pre}^*} Y ::= \mathrm{AStore}\ \langle R, T \rangle\ (X \rightsquigarrow_{\widehat{pre}} Y)$$

$$\begin{aligned}
&\mathrm{random}_{\widehat{pre}^*} : X \rightsquigarrow_{\widehat{pre}^*} [0,1] \\
&\mathrm{random}_{\widehat{pre}^*}\ j := \\
&\quad \mathrm{fst}_{\widehat{pre}} \ggg_{\widehat{pre}} \mathrm{fst}_{\widehat{pre}} \ggg_{\widehat{pre}} \pi_{\widehat{pre}}\ j
\end{aligned}$$

$$\begin{aligned}
&\mathrm{branch}_{\widehat{pre}^*} : X \rightsquigarrow_{\widehat{pre}^*} \mathrm{Bool} \\
&\mathrm{branch}_{\widehat{pre}^*}\ j := \\
&\quad \mathrm{fst}_{\widehat{pre}} \ggg_{\widehat{pre}} \mathrm{snd}_{\widehat{pre}} \ggg_{\widehat{pre}} \pi_{\widehat{pre}}\ j
\end{aligned}$$

$$\mathrm{fst}_{\widehat{pre}^*} := \eta_{\widehat{pre}^*}\ \mathrm{fst}_{\widehat{pre}};\ \cdots$$

$$\begin{aligned}
&\mathrm{ifte}^{\Downarrow}_{\widehat{pre}^*} : (X \rightsquigarrow_{\widehat{pre}^*} \mathrm{Bool}) \to (X \rightsquigarrow_{\widehat{pre}^*} Y) \to (X \rightsquigarrow_{\widehat{pre}^*} Y) \\
&\qquad \to (X \rightsquigarrow_{\widehat{pre}^*} Y) \\
&\mathrm{ifte}^{\Downarrow}_{\widehat{pre}^*}\ k_1\ k_2\ k_3\ j := \\
&\quad \mathrm{let}\ \langle C_k, p_k \rangle := k_1\ (\mathrm{left}\ j)\ A \\
&\qquad\quad \langle C_b, p_b \rangle := \mathrm{branch}_{\widehat{pre}^*}\ j\ A \\
&\qquad\qquad C_2 := C_k \cap C_b \cap \{\mathrm{true}\} \\
&\qquad\qquad C_3 := C_k \cap C_b \cap \{\mathrm{false}\} \\
&\qquad\qquad A_2 := p_k\ C_2 \cap p_b\ C_2 \\
&\qquad\qquad A_3 := p_k\ C_3 \cap p_b\ C_3 \\
&\quad \mathrm{in}\ \mathrm{if}\ C_b = \{\mathrm{true}, \mathrm{false}\} \\
&\qquad \mathrm{then}\ \langle Y, \lambda B.\, A_2 \sqcup A_3 \rangle \\
&\qquad \mathrm{else}\ k_2\ (\mathrm{left}\ (\mathrm{right}\ j))\ A_2 \cup_{\widehat{pre}} k_3\ (\mathrm{right}\ (\mathrm{right}\ j))\ A_3
\end{aligned}$$

(d) Abstract preimage* arrow combinators for probabilistic choice and guaranteed termination. Fig. 6 defines $\eta_{\widehat{pre}^*}$, $(\ggg_{\widehat{pre}^*})$, $(\&\&\&_{\widehat{pre}^*})$, $\mathrm{ifte}_{\widehat{pre}^*}$ and $\mathrm{lazy}_{\widehat{pre}^*}$.

Fig. 7: Implementable arrows that approximate preimage arrows.

Fortunately, implementing $[\![\cdot]\!]_{\widehat{\mathsf{pre}}}$ as defined in Fig. 2 requires lifting only a few pure functions: id, fst, snd, const $v$ for any literal constant $v$, and primitives $\delta$. According to (30) and (31), implementing the extended semantics $[\![\cdot]\!]^{\Downarrow}_{\widehat{\mathsf{pre}}^*}$, which supports random choice and guarantees termination, requires lifting only $\pi$ j for any $\mathsf{j} \in \mathsf{J}$. Fig. 7c gives explicit definitions for $\mathsf{id}_{\widehat{\mathsf{pre}}}$, $\mathsf{fst}_{\widehat{\mathsf{pre}}}$, $\mathsf{snd}_{\widehat{\mathsf{pre}}}$, $\mathsf{const}_{\widehat{\mathsf{pre}}}$ and $\pi_{\widehat{\mathsf{pre}}}$.

Fig. 7d defines the abstract preimage* arrow using the AStore arrow transformer (see Fig. 6), in terms of the abstract preimage arrow, and defines $\mathsf{random}_{\widehat{\mathsf{pre}}^*}$ and $\mathsf{branch}_{\widehat{\mathsf{pre}}^*}$ using the manual lifts in Fig. 7c.

Guaranteeing termination requires some care. The definition of $\mathsf{ifte}^{\Downarrow}_{\widehat{\mathsf{pre}}^*}$ in Fig. 7d is obtained by expanding the definition of $\mathsf{ifte}^{\Downarrow}_{\mathsf{pre}^*}$, and changing the case in which the set of branch traces allows both branches. Instead of taking both branches, it takes neither, and returns a loose but sound approximation.

### 6.3 Correctness and Termination

Let $\mathsf{h} := [\![e]\!]^{\Downarrow}_{\mathsf{pre}^*} : \mathsf{X} \rightsquigarrow_{\mathsf{pre}^*} \mathsf{Y}$ and $\widehat{\mathsf{h}} := [\![e]\!]^{\Downarrow}_{\widehat{\mathsf{pre}}^*} : \mathsf{X} \rightsquigarrow_{\widehat{\mathsf{pre}}^*} \mathsf{Y}$ for some expression $e$.

**Theorem 8 (terminating, monotone, sound and decreasing).** *For all* $\mathsf{A} : \mathsf{Rect}\ \langle\langle \mathsf{R}, \mathsf{T}\rangle, \mathsf{X}\rangle$ *and* $\mathsf{B} : \mathsf{Rect}\ \mathsf{Y}$,

- $\mathsf{ap}_{\widehat{\mathsf{pre}}}\ (\widehat{\mathsf{h}}\ \mathsf{j}_0\ \mathsf{A})\ \mathsf{B}$ *terminates.*
- $\lambda \mathsf{A}'.\, \mathsf{ap}_{\widehat{\mathsf{pre}}}\ (\widehat{\mathsf{h}}\ \mathsf{j}_0\ \mathsf{A}')\ \mathsf{B}$ *and* $\lambda \mathsf{B}'.\, \mathsf{ap}_{\widehat{\mathsf{pre}}}\ (\widehat{\mathsf{h}}\ \mathsf{j}_0\ \mathsf{A})\ \mathsf{B}'$ *are monotone.*
- $\mathsf{ap}_{\mathsf{pre}}\ (\mathsf{h}\ \mathsf{j}_0\ \mathsf{A})\ \mathsf{B}\ \subseteq\ \mathsf{ap}_{\widehat{\mathsf{pre}}}\ (\widehat{\mathsf{h}}\ \mathsf{j}_0\ \mathsf{A})\ \mathsf{B}\ \subseteq\ \mathsf{A}$ *(i.e. sound and decreasing).*

Given these properties, we might try to compute preimages of B by computing preimages restricted to the parts of increasingly fine discretizations of A.

**Definition 13 (preimage refinement algorithm).** *Let* $\mathsf{B} : \mathsf{Rect}\ \mathsf{Y}$. *Define*

$$\begin{aligned} &\mathsf{refine} : \mathsf{Rect}\ \langle\langle \mathsf{R}, \mathsf{T}\rangle, \mathsf{X}\rangle \to \mathsf{Rect}\ \langle\langle \mathsf{R}, \mathsf{T}\rangle, \mathsf{X}\rangle \\ &\mathsf{refine}\ \mathsf{A}\ :=\ \mathsf{ap}_{\widehat{\mathsf{pre}}}\ (\widehat{\mathsf{h}}\ \mathsf{j}_0\ \mathsf{A})\ \mathsf{B} \end{aligned} \tag{41}$$

*Define* $\mathsf{partition} : \mathsf{Rect}\ \langle\langle \mathsf{R}, \mathsf{T}\rangle, \mathsf{X}\rangle \to \mathsf{Set}\ (\mathsf{Rect}\ \langle\langle \mathsf{R}, \mathsf{T}\rangle, \mathsf{X}\rangle)$ *to produce positive-measure, disjoint rectangles, and define*

$$\begin{aligned} &\mathsf{refine}^* : \mathsf{Set}\ (\mathsf{Rect}\ \langle\langle \mathsf{R}, \mathsf{T}\rangle, \mathsf{X}\rangle) \to \mathsf{Set}\ (\mathsf{Rect}\ \langle\langle \mathsf{R}, \mathsf{T}\rangle, \mathsf{X}\rangle) \\ &\mathsf{refine}^*\ \mathcal{A}\ :=\ \mathsf{image\ refine}\ \left(\textstyle\bigcup_{\mathsf{A} \in \mathcal{A}} \mathsf{partition}\ \mathsf{A}\right) \end{aligned} \tag{42}$$

*For any* $\mathsf{A} : \mathsf{Rect}\ \langle\langle \mathsf{R}, \mathsf{T}\rangle, \mathsf{X}\rangle$, *iterate* $\mathsf{refine}^*$ *on* $\{\mathsf{A}\}$.

Monotonicity ensures refining a partition of A never does worse than refining A itself, decreasingness ensures $\mathsf{refine}\ \mathsf{A} \subseteq \mathsf{A}$, and soundness ensures the preimage of B is covered by the partition $\mathsf{refine}^*$ returns. Ideally, the algorithm would be complete, in that covering partitions converge to a set that overapproximates by a measure-zero subset. Unfortunately, convergence fails on some examples that terminate with probability less than one. We leave completeness conditions for future work, and for now, use algorithms that depend only on soundness.

## 7 Implementations and Examples

This section describes our implementations and gives examples, including probabilistic verification of floating-point error bounds.

We have three implementations: two direct implementations of the abstract semantics, and a less direct but more efficient one called **Dr. Bayes**. All of them can be found at `https://github.com/ntoronto/drbayes`.

Given a library for operating on rectangular sets, the abstract preimage arrows defined in Figs. 6 and 7 can be implemented with few changes in any practical $\lambda$-calculus. We have done so in Typed Racket [27] and Haskell [1]. Both implementations are almost line-for-line transliterations from the figures.

Dr. Bayes is written in Typed Racket. It includes $[\![\cdot]\!]_{\mathsf{a}^*}$ (Fig. 2), its extension $[\![\cdot]\!]^{\Downarrow}_{\mathsf{a}^*}$, the bottom* arrow (Figs. 3 and 6), the abstract preimage and preimage* arrows (Figs. 7 and 6), and other manual lifts to compute abstract preimages under real functions such as arithmetic, `sqrt` and `log`. The abstract preimage arrows operate on a monomorphic rectangular set data type, which includes tagged rectangles and disjoint unions for ad-hoc polymorphism, and floating-point intervals to overapproximate real intervals.

Definition 13 outlines preimage refinement, a discretization algorithm that repeatedly shrinks and repartitions a program's domain. *Dr. Bayes does not use this algorithm directly* because it is inefficient: good accuracy requires fine discretization, which is exponential in the number of discretized axes. Instead of *enumerating* covering partitions of the random source, Dr. Bayes *samples parts* from the covering partitions and then *samples a point* from each sampled part, with time complexity linear in the number of samples and discretized axes. It applies bottom* arrow computations to the random source samples to get output samples, rejecting those outside the requested output set.

In short, Dr. Bayes uses preimage refinement only to reduce the rate of rejection when sampling under constraints, and thus relies only on its soundness.

We have tested Dr. Bayes on a variety of Bayesian inference tasks, including Bayesian regression and model selection [28]. Some of our Bayesian inference tests use recursion and constrain the outputs of deterministic functions, suggesting that Dr. Bayes and future probabilistic languages like it will allow practitioners to model real-world processes more expressively and precisely.

Recent work in probabilistic verification recasts it as a probabilistic inference task [9]. Given that Dr. Bayes's runtime is designed to sample efficiently under low-probability constraints, using it to probabilistically verify that a program does not exhibit certain errors is fairly natural. To do so, we

1. Encode the program in a way that propagates and returns errors.
2. Run the program with the constraint that the output is an error.

Sometimes, Dr. Bayes can determine that the preimage of the constrained output set is $\varnothing$, which is a proof that the program never exhibits an error. Otherwise, the longer the program runs without returning samples, the likelier it is that the preimage has zero probability or is empty; i.e. that an error does not occur.

As an extended example, we consider verifying floating-point error bounds.

While Dr. Bayes's numbers are implemented by floating-point intervals, semantically, they are real numbers. We therefore cannot easily represent floating-point numbers in Dr. Bayes—but we do not want to. We want *abstract* floating-point numbers, each consisting of an exact, real number and a bound on the relative error with which it is approximated. We define the following two structures to represent abstract floats.

```
(struct/drbayes float-any ())
(struct/drbayes float (value error))
```

An abstract value `(float v e)` represents every float between `(* v (- 1 e))` and `(* v (+ 1 e))` inclusive, while `(float-any)` represents NaN and other catastrophic error conditions. Abstract floating-point functions such as `flsqrt` compute exact results and use input error to compute bounds on output error:

```
(define/drbayes (flsqrt x)
  (if (float-any? x)
      x
      (let ([v  (float-value x)]
            [e  (float-error x)])
        (cond [(negative? v)  (float-any)]  ; NaN
              [(zero? v)      (float 0 0)]  ; exact case
              [else  ; v is positive
               (float (sqrt v)                    ; exact square root
                      (+ (- 1 (sqrt (- 1 e)))     ; relative error
                         (* 1/2 epsilon)))]))))  ; rounding error
```

We have similarly implemented abstract floating-point arithmetic, comparison, exponentials, and logarithms in Dr. Bayes.

Suppose we define an abstract floating-point implementation of the geometric distribution's inverse CDF using the formula $(\mathsf{log}\ \mathsf{u})/(\mathsf{log}\ (1-\mathsf{p}))$:

```
(define/drbayes (flgeometric-inv-cdf u p)
  (fl/ (fllog u) (fllog (fl- (float 1 0) p))))
```

We want the distribution of $\langle \mathsf{u}, \mathsf{p} \rangle$ in $(0,1) \times (0,1)$ with the value of

```
(float-error (flgeometric-inv-cdf (float u 0) (float p 0)))
```

constrained to $(3 \cdot \varepsilon, \infty)$, where $\varepsilon \approx 2.22 \cdot 10^{-16}$ is floating-point epsilon for 64-bit floats. That is, we want the distribution of inputs for which the floating-point output may be more than 3 epsilons away from the exact output.

Dr. Bayes returns samples of $\langle \mathsf{u}, \mathsf{p} \rangle$ within about $(0,1) \times (\varepsilon, 0.284)$, a fairly large domain on which error is greater than 3 epsilons. Realizing that the rounding error in $1 - \mathsf{p}$ is magnified by $\mathsf{log}$'s relative error when $\mathsf{p}$ is small, we define

```
(define/drbayes (flgeometric-inv-cdf u p)
  (fl/ (fllog u) (fllog1p (flneg p))))
```

where `fllog1p` (abstractly) computes $\mathsf{log}\ (1 + \mathsf{x})$ with high accuracy. Dr. Bayes reports that the preimage of $(3 \cdot \varepsilon, \infty)$ is $\varnothing$. In fact, the preimage of $(1.51 \cdot \varepsilon, \infty)$ is $\varnothing$, so `flgeometric-inv-cdf` introduces error of no more than 1.51 epsilons.

We have used this technique to verify error bounds on the implementations of hypot, sqrt1pm1 and sinh in Racket's `math` library.

## 8 Related Work

Probabilistic languages can be approximately placed into two groups: those defined by a semantics, and those defined by an implementation.

Kozen's seminal work [15] on probabilistic semantics defines two measure-theoretic, denotational semantics, in two different styles: a **random-world semantics** [18] that interprets programs as deterministic functions that operate on a random source, and a **distributional semantics** that interprets programs as probability measures. It seems that all semantics work thereafter is in one of these styles. For example, Hurd [11] develops a random-world semantics in HOL and uses it to formally verify randomized algorithms such as the Miller-Rabin primality test. Ours is also a random-world semantics.

Jones [12] defines the probability monad as a categorical metatheory for interpreting probabilistic programs as distributions. Ramsey and Pfeffer [25] reformulate it in terms of Haskell's `return` and '$\gg=$', and use it to define a distributional semantics for a probabilistic lambda calculus. They implement the probability monad using probability mass functions, show that computing certain queries is inefficient, and devise an equivalent semantics that is more amenable to efficient implementation, for programs with finite probabilistic choice.

To put Infer.NET [21] on solid footing, Borgström et al. [4] define a distributional semantics for a first-order probabilistic language with bounded loops and constraints, by transforming terms into arrow-like combinators that produce measures. But Infer.NET interprets programs as probability density functions,[4] so they develop a semantics that does the same and prove equivalence.

The work of Borgström et al. and Ramsey and Pfeffer exemplify a larger trend: while *defining* probabilistic languages can be done using measure theory, *implementing* them to support more than just evaluation (such as allowing constraints) has seemed hopeless enough to necessitate using a less explanatory theory of probability that has more obvious computational content. Indeed, the distributional semantics of Pfeffer's IBAL [24] and Nori et al.'s R2 [23] are defined in terms of probability mass and density functions in the first place. R2 lifts some of the resulting restrictions and speeds up sampling by propagating constraints toward the random values they refer to.

Some languages defined by an implementation are probabilistic Scheme [14], BUGS [17], BLOG [20], BLAISE [3], Church [8], and Kiselyov's embedded language for OCaml [13]. Recently, Wingate et al. [32] define nonstandard semantics that enable efficient inference, but do not define the languages. All of these languages are implemented in terms of probability mass or density functions.

Our work is similar in structure to monadic abstract interpretation [26,6], which also parameterizes a semantics on categorical meanings.

Cousot's probabilistic abstract interpretation [5] is a general framework for static analyses of probabilistic languages. It considers only random-world semantics, which is quite practical: because programs are interpreted as deterministic functions, many existing analyses easily apply. Our random-world semantics fits in this framework, but the concrete domain of preimage functions does not appear among Cousot's many examples, and we do not compute fixed points.

[4] More precisely, as factor graphs, which represent probability density functions.

## 9 Conclusions and Future Work

To allow arbitrary constraints and recursion in probabilistic programs, we combined the power of measure theory with the unifying elegance of arrows. We (a) defined a transformation from first-order programs to arbitrary arrows, (b) defined the bottom arrow as a standard translation target, (c) derived the uncomputable preimage arrow as an alternative target, and (d) derived a sound, computable approximation of the preimage arrow, and enough computable lifts to transform programs. We implemented the abstract semantics and carried out Bayesian inference, stochastic ray tracing, and probabilistic verification.

In the future, we intend to add expressiveness by adding lambdas (possibly via closure conversion), explore ways to use static or dynamic analyses to speed up Monte Carlo algorithms, and explore preimage computation's connections to type checking and type inference. More broadly, we hope to advance probabilistic inference by providing a rich modeling language with an efficient, correct implementation, which allows general recursion and arbitrary constraints.

**Acknowledgments.** Special thanks to Mitchell Wand for careful review and helpful feedback. This material is based on research sponsored by DARPA under the Automated Program Analysis for Cybersecurity (FA8750-12-2-0106) project. The U.S. Government is authorized to reproduce and distribute reprints for Governmental purposes notwithstanding any copyright notation thereon.